\documentclass[preprint,11pt,nofootinbib]{revtex4-1}

\usepackage{graphicx}
\usepackage{dcolumn} 
\usepackage{bm}
\usepackage{amssymb}
\usepackage{amsmath}
\usepackage{epsfig}    
\usepackage{color}
\usepackage{slashed}
\usepackage{hyperref}

\begin{document} 

\title{Dark Sector as Origin of Light Lepton Mass and Its Phenomenology}
\preprint{OU-HET-1127}

\author{Cheng-Wei Chiang}
\email{chengwei@phys.ntu.edu.tw}
\affiliation{Department of Physics and Center for Theoretical Physics, National Taiwan University, Taipei, Taiwan 10617, R.O.C.}
\affiliation{Physics Division, National Center for Theoretical Sciences, Taipei, Taiwan 10617, R.O.C.}

\author{Ryomei Obuchi}
\email{r\_obuchi@hetmail.phys.sci.osaka-u.ac.jp}
\affiliation{Department of Physics, Osaka University, Toyonaka, Osaka 560-0043, Japan}

\author{Kei Yagyu}
\email{yagyu@het.phys.sci.osaka-u.ac.jp}
\affiliation{Department of Physics, Osaka University, Toyonaka, Osaka 560-0043, Japan}

\begin{abstract}

We discuss a model with a dark sector, in which smallness of mass for charged leptons and neutrinos can naturally be explained by one-loop effects mediated by particles in the dark sector.  These new particles, including dark matter candidates, also contribute to the anomalous magnetic dipole moment, denoted by $(g-2)$, for charged leptons.  We show that our model can explain the muon $(g-2)$ anomaly and observed neutrino oscillations under the constraints from lepton flavor violating decays of charged leptons.  We illustrate that the scenario with scalar dark matter is highly constrained by direct searches at the LHC, while that with fermionic dark matter allows for considering dark scalars with masses of order 100~GeV.  Our scenario can be tested by a precise measurement of the muon Yukawa coupling as well as the 
direct production of dark scalar bosons at future electron-positron colliders. 

\end{abstract}
\maketitle

\section{Introduction}

The Higgs mechanism that generates mass for weak gauge bosons through the spontaneous breakdown of the electroweak symmetry has been verified by the discovery of the Higgs boson at LHC.  In addition, mass generation for charged fermions through the Yukawa interaction in the Standard Model (SM) is so far consistent with the observed Higgs decays into bottom quarks~\cite{Sirunyan:2018kst,Aaboud:2018zhk} and tau leptons~\cite{Aad:2015vsa,Sirunyan:2017khh} at the CERN LHC. 
Furthermore, the Higgs to dimuon channel has also been explored, and its observed significance has currently been measured to be 2.0$\sigma$~\cite{Aad:2020xfq} and 3.0$\sigma$~\cite{Sirunyan:2020two} at the ATLAS and CMS experiments, respectively. 
 This, however, does not mean that all the known fermion masses are necessarily generated by the same mechanism, especially when a large hierarchy exists among various Yukawa couplings.  In particular, mass of charged fermions in the first two generations and neutrinos can be generated through other means, rather than the SM Yukawa interaction, because of the somewhat unnatural small Yukawa couplings.

Such tiny masses can naturally be explained by forbidding tree-level Yukawa couplings and introducing quantum effects from what is referred to as a dark sector.  In such a scenario, the lightest neutral particle in the dark sector can serve as a dark matter candidate.  In addition, the anomaly of the muon anomalous magnetic dipole moment, $(g-2)_\mu$, of about 4.2~$\sigma$ discrepancy between the value observed at Fermilab and the SM prediction~\cite{Abi2021}, can be explained by loop effects of the dark sector particles.  Interestingly, it has been shown in Ref.~\cite{Chiang:2021pma} that the new contribution to the muon $(g-2)$ does not explicitly depend on new couplings among particles in the SM and dark sector, but is essentially determined by the mass of a vector-like lepton in the dark sector.  This is simply because one-loop diagrams for the muon mass generation and $(g-2)$ are realized by the common loop effect of the dark sector.  Therefore, in this scenario, small masses for charged leptons, dark matter and the muon $(g-2)$ can be simultaneously explained through a common origin of the dark sector.  

In this paper, we discuss phenomenological consequences of the model with a dark sector as proposed in Refs.~\cite{Chiang:2021pma}
\footnote{See also Refs.~\cite{Ma:2013mga,Baker:2021yli,Gabrielli:2016vbb,Gabrielli:2013jka} for the radiative generation of masses of charged fermions.},
where masses for electrons and muons are generated at the one-loop level.  However, the mechanism for neutrino mass generation was not elucidated in the model.  It is one object of this work to extend the model so as to accommodate the radiative mass generation for neutrinos as well as electron and muon.  We find that it can be realized by adding right-handed neutrinos $\nu_R^{}$ to the dark sector.  In order to make phenomenologically acceptable scenarios, we impose a key assumption that the lepton flavor and lepton number symmetries are explicitly broken only via Majorana mass for $\nu_R^{}$.  We can then successfully accommodate observed neutrino oscillations without contradiction to data for lepton flavor violating (LFV) decays of charged leptons.  We show that such an extension is not only required for the neutrino mass generation, but also to avoid severe constraints from direct searches for 
new particles charged under the electroweak symmetry, such as the electroweakinos in supersymmetric (SUSY) models at the LHC~\cite{ATLAS:2021moa}.  In our model with $\nu_R^{}$, the main decay modes of the dark scalars can be replaced by those via new Yukawa couplings with $\nu_R^{}$, {\it e.g.}, the neutral scalars can decay into $\bar{\nu}_L^{}\nu_R^{}$ which escape LHC detectors.  With light dark scalar particles of mass about 100~GeV that are allowed by the current LHC data, we find successful benchmark scenarios that can explain the muon $(g-2)$ anomaly, neutrino oscillations and data of LFV decays for charged leptons.  These light new particles can directly be produced at future lepton colliders such as the International Linear Collider (ILC)~\cite{Baer:2013cma,Asai:2017pwp,Fujii:2017vwa}, the Circular Electron Positron Collider (CEPC)~\cite{CEPC-SPPCStudyGroup:2015csa} and the Future Circular Collider (FCC-ee)~\cite{Gomez-Ceballos:2013zzn}.

This paper is organized as follows.  In Sec.~\ref{sec:model}, we define the model with a dark sector, including the new particles and their quantum numbers.  In Sec.~\ref{sec:lepton}, we discuss radiative mass generation for charged leptons and neutrinos, and consider new contributions to the muon $(g-2)$.  We also take into account the constraints from LFV decays of charged leptons.  Sec.~\ref{sec:collider} is devoted to collider phenomenology.  We first consider the current direct search bounds from the LHC.  We then discuss the deviation in the muon Yukawa coupling and direct productions for dark scalar bosons at the future lepton colliders.  Conclusions are drawn in Sec.~\ref{sec:conclusions}.

\section{Model \label{sec:model}}

We first briefly review the model with a dark sector proposed in Ref.~\cite{Chiang:2021pma}, where the masses of electron and muon are generated at one-loop level.  The dark sector is defined by introducing an unbroken $Z_2$ symmetry, with all particles in the dark sector (SM particles) being assigned to be $Z_2$ odd (even).  The lightest $Z_2$ odd particle would then be a dark matter candidate.  In addition, we introduce a softly-broken $Z_2'$ symmetry in order to forbid tree level mass for charged leptons, particularly the electron and muon.  Furthermore, a global $U(1)_\ell$ symmetry is imposed on the model to suppress LFV decays of charged leptons (such as $\mu \to e \gamma$) and has to be softly-broken for neutrino mass generation.  In short, our model has the symmetry $SU(2)_I\times U(1)_Y \times U(1)_\ell \times Z_2 \times Z_2'$, where $SU(2)_I\times U(1)_Y$ is the electroweak gauge symmetry.

\begin{table}[t]
\begin{center}
\begin{tabular}{|c||cccc|ccc|}\hline
             & \multicolumn{4}{c|}{Fermions}  &  \multicolumn{3}{c|}{Scalars} \\\hline\hline
 Fields      &  $(L_L^e,L_L^\mu,L_L^\tau)$       & $(e_R,\mu_R^{},\tau_R^{})$     & $(F_{L/R}^e,F_{L/R}^\mu)$  & $(\nu_R^e,\nu_R^\mu,\nu_R^\tau)$  & $H$     & $\eta^{}$ & $ S^+$  \\\hline\hline
$SU(2)_I$    &  ${\bm 2}$              & ${\bm 1}$         & ${\bm 1}$                & ${\bm 1}$                        & ${\bm 2}$ & ${\bm 2}$ & ${\bm 1}$   \\\hline
$U(1)_Y$     &  $-1/2$                 & $-1$              & $0$                      & $0$                              & $1/2$ & $1/2$  & $1$\\\hline
$U(1)_\ell$  & $(q_e,q_\mu,q_\tau)$            &  $(q_e,q_\mu,q_\tau)$    & $(q_e,q_\mu)$             & $(q_e,q_\mu,q_\tau)$                  & $0$&$0$& $0$ \\\hline
$Z_2$        & $+$                 &  $+$          & $-$                  & $-$                              & $+$ & $-$& $-$  \\\hline
$Z_2'$       & $+$                 &  $(-,-,+)$          & $+/-$              & $-$                               & $+$& $-$&$-$\\\hline
\end{tabular}
\caption{Particle contents and charge assignments under the $SU(2)_I\times U(1)_Y \times U(1)_\ell \times Z_2 \times Z_2'$ symmetry.  }
\label{particle}
\end{center}
\end{table}

Although there are several concrete scenarios for this model~\cite{Chiang:2021pma}, we focus exclusively on the simplest one~\cite{Ma:2013mga} whose dark sector is composed of an isospin doublet scalar $\eta$, a pair of charged singlet scalars $S^\pm$ and singlet vector-like leptons $F^{e,\mu}$.  We now further introduce three right-handed neutrinos $\nu_R^{i}$ for the purpose of generating Majorana mass for left-handed neutrinos while realizing a phenomenologically acceptable scenario, which will be discussed later.  In Table~\ref{particle}, we summarize the particle contents and charge assignments under the symmetry explained above.

The Higgs doublet field $H$ and the dark doublet field $\eta$, whose neutral component is assumed not to develop a vacuum expectation value (VEV), can be parameterized as follows: 
\begin{align}
H = \begin{pmatrix}
G^+ \\
\frac{h + v + i G^0}{\sqrt{2}}
\end{pmatrix} 
~,\quad 
\eta  = 
\begin{pmatrix}
\eta^+ \\
\frac{\eta_R^{} + i \eta_I^{}}{\sqrt{2}}
\end{pmatrix}
~, 
\end{align}
where the VEV $v = (\sqrt{2}G_F)^{-1/2} = 246$~GeV with $G_F$ being the Fermi decay constant, $G^\pm$ ($G^0$) are the Nambu-Goldstone (NG) bosons absorbed into the longitudinal components of $W^\pm$ ($Z$), and $h$ is identified with the 125-GeV Higgs boson.  We note that the charged components $\eta^\pm$ can mix with $S^\pm$ to be discussed below.

The Yukawa interactions for leptons and the mass for $F^\ell$ are given by
\begin{align}
\begin{split}
{\cal L}_{\rm lep} &= - y_\tau\overline{L_L^\tau} H\tau_R  
 - \sum_{\ell=e,\mu} \left(M_\ell \overline{F_L^\ell} F_R^\ell + f_L^\ell \overline{L_L^\ell} \eta^c F_R^\ell + f_R^\ell \overline{\ell_R}S^- F_L^\ell \right)
- \sum_{\ell =e,\mu,\tau} \tilde{y}_R^\ell \overline{L_L^\ell} \eta^c \nu_R^\ell  + \text{H.c.} 
~, 
\end{split}
\label{eq:yukawa}
\end{align}
where $\eta^c \equiv i\tau_2\eta^*$ with $\tau_2$ being the second Pauli matrix and $S^- = (S^+)^*$.  As mentioned above, the Yukawa couplings for electrons and muons are absent at tree level due to the $Z_2'$ symmetry.  The new couplings $f_L^\ell$ and $f_R^\ell$ ($\tilde{y}_R^\ell$) are necessary for one-loop generation of the masses of electron and muon (left-handed neutrinos), to be discussed in more detail in the next section.  Due to the $U(1)_{\ell}$ invariance, the summations above is in a flavor-diagonal fashion, and the interaction $\overline{\ell_R}S^-\nu_R^{c}$ is thus forbidden.

If $U(1)_{\ell}$ is exact, the lepton flavors and the lepton number are conserved as in the SM.  As a result, both right-handed neutrinos $\nu_R^{}$ and left-handed neutrinos $\nu_L^{}$ should be massless. \footnote{The Dirac mass term $\overline{L_L} H^c \nu_R^{}$ is forbidden by the $Z_2$ symmetry. }  In order to make the neutrinos massive, we assume that the $U(1)_\ell$ symmetry is softly broken only via the Majorana mass term for $\nu_R^{}$: 
\begin{align}
\Delta{\cal L}_{\rm lep} &= 
- \sum_{\ell,\ell'=e,\mu,\tau}\frac{(\tilde{M}_R)_{\ell\ell'}}{2}\overline{\nu_R^{\ell c}}\nu_R^{\ell'} + \text{H.c.} 
\label{eq:yukawa2}
\end{align}
In general, $\tilde{M}_R$ is an arbitrary $3\times 3$ complex matrix, and we will take a basis transformation of $\nu_R^{}$ such that the mass matrix $\tilde{M}_R$ takes a diagonal form.  In such a physical basis, terms for the right-handed neutrinos are rewritten as 
\begin{align}
{\cal L}_{\nu_R} &= - \sum_{i=1,2,3}\frac{M_R^i}{2}\overline{\nu_R^{i c}}\nu_R^{i}  
- \sum_{\ell = e,\mu,\tau}\sum_{i = 1,2,3}y_R^{\ell i}\, \overline{L_L^\ell} \eta^c \nu_R^i  + \text{H.c.}
\label{eq:yukawa3}
\end{align}
In this basis, the Yukawa coupling $y_R$ is a general complex $3\times 3$ matrix.  For simplicity, we assume that all the parameters in the lepton sector are taken to be real. 

The most general scalar potential is given by 
\begin{align} 
V 
=& 
-\mu_H^2|H|^2 + \mu_\eta^2|\eta|^2 +  \mu_S^2|S^+|^2   + \left[\kappa \eta^T(i\tau_2) H S^- + \text{H.c.}\right] \notag\\
& 
+ \frac{\lambda_1}{2} |H|^4 + \frac{\lambda_2}{2} |\eta|^4 
  + \lambda_3 |H|^2|\eta|^2 + \lambda_4 |H^\dagger \eta|^2  + \left[\frac{\lambda_5}{2} \left(H^\dagger \eta \right)^2 + \text{H.c.} \right]\notag\\
& 
+ \frac{\lambda_6}{2}|S^+|^4  + \lambda_7 |H|^2|S^+|^2  + \lambda_8 |\eta|^2|S^+|^2. 
\end{align}
By rephasing the scalar fields, all the parameters in the above potential can be taken to be real without loss of generality.  The $\kappa$ term gives rise to a mixing between $\eta^\pm$ and $S^\pm$ as alluded to earlier, so that we 
define their mass eigenstates as follows: 
\begin{align}
\begin{pmatrix}
\eta^\pm\\
S^\pm
\end{pmatrix}
= 
\begin{pmatrix}
c_\theta & -s_\theta \\
s_\theta &  c_\theta 
\end{pmatrix}
\begin{pmatrix}
\eta_1^\pm\\
\eta_2^\pm
\end{pmatrix}
~, 
\end{align}
with the shorthand notation $c_X^{} = \cos X$ and $s_X^{} = \sin X$ and the mixing angle $\theta$ satisfying
\begin{align}
\tan2\theta &= \frac{2({\cal M}_\pm^2)_{12}}{({\cal M}_\pm^2)_{11}-({\cal M}_\pm^2)_{22}}
~. \label{eq:mixing}
\end{align}
In Eq.~(\ref{eq:mixing}), $({\cal M}_\pm^2)_{ij}$ are the elements of the squared mass matrix in the basis of $(\eta^{\pm},S^{\pm})$ given by
\begin{align}
{\cal M}_\pm^2 = \begin{pmatrix}
\mu_\eta^2 + \frac{v^2}{2}\lambda_3 & \frac{v\kappa}{\sqrt{2}} \\
\frac{v\kappa}{\sqrt{2}}& \mu_S^2 + \frac{v^2}{2} \lambda_7
\end{pmatrix}
~. 
\end{align}
The squared masses of the dark scalar bosons are given by
\begin{align}
\begin{split}
m_{\eta_1^\pm}^2 &= c_\theta^2 ({\cal M}_\pm^2)_{11} + s_\theta^2 ({\cal M}_\pm^2)_{22} + s_{2\theta} ({\cal M}_\pm^2)_{12}
~, \\
m_{\eta_2^\pm}^2 &= s_\theta^2 ({\cal M}_\pm^2)_{11} + c_\theta^2 ({\cal M}_\pm^2)_{22} - s_{2\theta} ({\cal M}_\pm^2)_{12}
~, \\
m_{\eta_R}^2  & = \mu_\eta^2 + \frac{v^2}{2}(\lambda_3  +\lambda_4  + \lambda_5)
~, \\
m_{\eta_I}^2  & = \mu_\eta^2 + \frac{v^2}{2}(\lambda_3  + \lambda_4 - \lambda_5)
~. 
\end{split} 
\end{align}
On the other hand, the squared mass of $h$ is given in the way as the SM after imposing the tadpole condition: 
\begin{align}
m_h^2 = v^2\lambda_1
~. 
\end{align}
Now, we can choose the 12 independent parameters in the potential as: 
\begin{align}
v,~m_h,~~m_{\eta_1^\pm},~~m_{\eta_2^\pm},~~m_{\eta_R},~~m_{\eta_I},~~\theta,~~\lambda_{3,7}
~, \label{eq:param}
\end{align}
and the three parameters $\lambda_{2,6,8}$ for the quartic interactions among the dark scalars that do not appear in the mass formulae.  In terms of the parameters given in Eq.~(\ref{eq:param}), some of the parameters in the potential are rewritten as
\begin{align}
\begin{split}
\mu_\eta^2 & = m_{\eta_1^\pm}^2c_\theta^2 + m_{\eta_2^\pm}^2s_\theta^2 - \frac{v^2}{2}\lambda_3
~, \\
\mu_S^2 & = m_{\eta_1^\pm}^2s_\theta^2 + m_{\eta_2^\pm}^2c_\theta^2 - \frac{v^2}{2}\lambda_7
~, \\
\kappa & = \frac{\sqrt{2}}{v}s_\theta c_\theta(m_{\eta_1^\pm}^2 - m_{\eta_2^\pm}^2)
~, \\
\lambda_4 & = \frac{1}{v^2}(m_{\eta_R}^2 + m_{\eta_I}^2 - 2m_{\eta_1^\pm}^2 c_\theta^2 - 2m_{\eta_2^\pm}^2 s_\theta^2)
~, \\
\lambda_5 & = \frac{1}{v^2}(m_{\eta_R}^2 - m_{\eta_I}^2)
~. 
\end{split}
\end{align}

The parameters of the potential can be constrained by imposing the bounds from perturbative unitarity and vacuum stability.  For the former, we require that the magnitudes of eigenvalues of the $s$-wave amplitude matrix elements for elastic 2-to-2 scalar scatterings at tree level be smaller than 1/2.  In the high-energy limit, these conditions can be expressed as~\cite{Muhlleitner:2016mzt,Chen:2020tfr} 
\begin{align}
\begin{split}
&\left|\lambda_{1}+\lambda_{2}+\sqrt{\left(\lambda_{1}-\lambda_{2}\right)^{2}+4 \lambda_{4}^{2}} \right| <16\pi
~, \quad
\left|\lambda_{1}+\lambda_{2}+\sqrt{\left(\lambda_{1}-\lambda_{2}\right)^{2}+4 \lambda_{5}^{2}}\right|  <16\pi
~, \\
&\left|\lambda_3 + 2 \lambda_4 \pm 3 \lambda_5\right|<8 \pi
~, \quad
\left|\lambda_3 \pm \lambda_5\right| <8 \pi
~, \quad
\left|\lambda_3 \pm \lambda_4\right| <8 \pi
~, \quad 
\left|\lambda_{7,8}\right| <8 \pi
~,\quad 
\left|a_{1,2,3}\right| <8 \pi
~, \label{eq:per-lam268}
\end{split}
\end{align}
where $a_{1,2,3}$ are the eigenvalues of the following $3\times 3$ matrix
\begin{align}
\begin{pmatrix}
3\lambda_1 & 2\lambda_3 + \lambda_4 & \sqrt{2}\lambda_7 \\
2\lambda_3 + \lambda_4 & 3\lambda_2 & \sqrt{2}\lambda_8\\
\sqrt{2}\lambda_7  & \sqrt{2}\lambda_8  & 2\lambda_6
\end{pmatrix}
~. 
\end{align}

For the vacuum stability bound, we require that the potential be bounded from below in any direction at large scalar field values.  This is ensured by imposing the following conditions:~\cite{Muhlleitner:2016mzt}  
\begin{equation}
\lambda_i \in \Omega_1\cup\Omega_2,\quad i = 1, \ldots, 8
\end{equation}
where 
\begin{align}
&
\Omega_{1}=\Big\{\lambda_{1}, \lambda_{2}, \lambda_{6}>0 ; \sqrt{\lambda_{1} \lambda_{6}}+\lambda_{7}>0 ; \sqrt{\lambda_{2} \lambda_{6}}+\lambda_{8}>0 ; \notag \\
& \qquad\qquad
\sqrt{\lambda_{1} \lambda_{2}}+\lambda_{3}+D>0 ; \lambda_{7}+\sqrt{\frac{\lambda_{1}}{\lambda_{2}}} \lambda_{8} \geq 0\Big\}
~,
\label{eq:stability1}
\\
&
\Omega_{2}=\Big\{\lambda_{1}, \lambda_{2}, \lambda_{6}>0 ; \sqrt{\lambda_{2} \lambda_{6}} \geq \lambda_{8}>-\sqrt{\lambda_{2} \lambda_{6}} ; \sqrt{\lambda_{1} \lambda_{6}}>-\lambda_{7} \geq \sqrt{\frac{\lambda_{1}}{\lambda_{2}}} \lambda_{8} ; \notag \\
& \qquad\qquad
\sqrt{\left(\lambda_{7}^{2}-\lambda_{1} \lambda_{6}\right)\left(\lambda_{8}^{2}-\lambda_{2} \lambda_{6}\right)}>\lambda_{7} \lambda_{8}-\left(D+\lambda_{3}\right) \lambda_{6}\Big\}
~, 
\label{eq:stability2}
\end{align}
with $D = \max\left\{0, \lambda_4 - \lambda_5 \right\}$. 

In addition to these constraints, the masses and the mixing angle $\theta$ are constrained by taking into account the electroweak oblique parameters which are 
conveniently parameterized by the $S$, $T$ and $U$ parameters~\cite{Peskin:1990zt}. In particular, the $T$ parameter constrains mass differences among the dark scalar bosons. 
The new contribution to the $T$ parameter, $\Delta T$, is calculated as 
\begin{align}  
\Delta T =&~ \frac{\sqrt{2}G_F}{16\pi^2\alpha_{\text{em}}}\Bigg\{c_\theta^2 \left[F_T(m_{\eta_R^{}},m_{\eta_1^\pm}) + F_T(m_{\eta_I^{}},m_{\eta_1^\pm})\right] \notag\\
&+ s_\theta^2 \left[F_T(m_{\eta_R^{}},m_{\eta_2^\pm}) + F_T(m_{\eta_I^{}},m_{\eta_2^\pm})\right] -F_T(m_{\eta_R^{}},m_{\eta_I^{}})  \Bigg\}
~, 
\label{eq:deltat}
\end{align}
where $\alpha_{\text{em}}$ is the fine structure constant, and 
\begin{align}  
F_T(m_1,m_2) = \frac{1}{2}(m_1^2 + m_2^2) - \frac{m_1^2 m_2^2}{m_1^2 - m_2^2}\ln \frac{m_1^2}{m_2^2}
~. 
\end{align}
This function returns zero for $m_1 = m_2$.  If we take the no mixing limit, {\it i.e.}, $\theta\to 0$ ($\theta\to \pi/2$), $\Delta T$ vanishes for $m_{\eta_R^{}} = m_{\eta_1^\pm}\,(m_{\eta_2^\pm})$ or $m_{\eta_I^{}} = m_{\eta_1^\pm}\, (m_{\eta_2^\pm})$, because 
the $\eta_1^\pm$ ($\eta_2^\pm$) states coincide with $\eta^\pm$ and then the custodial $SU(2)_L\times SU(2)_R$ symmetry is restored~\cite{Pomarol:1993mu}.

\section{Lepton Masses, $(g-2)$ and LFV decays \label{sec:lepton}}

\begin{figure}[t]
\begin{center}
 \includegraphics[width=150mm]{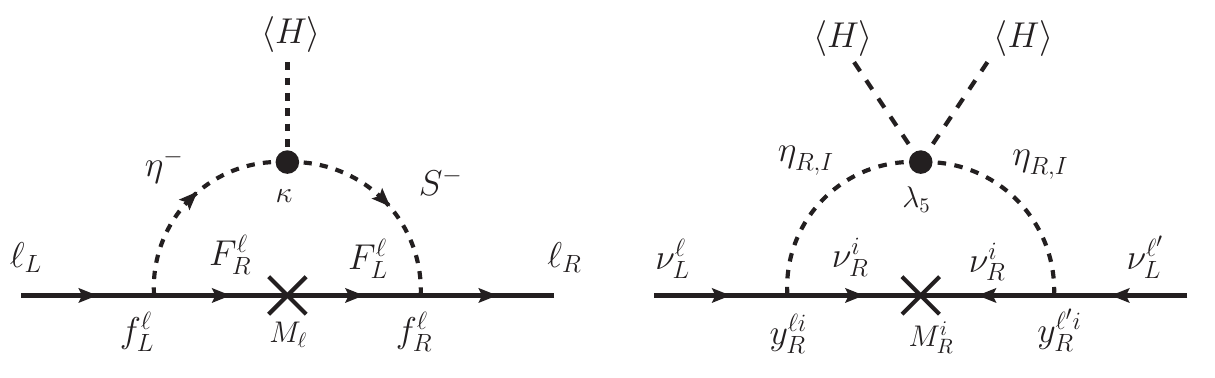}
   \caption{One-loop diagram for generating the mass for charged leptons (left) and left-handed neutrinos (right). }
   \label{fig:diagram}
\end{center}
\end{figure}

In our model, the masses of electron and muon are generated at the one-loop level as shown in the left plot of Fig.~\ref{fig:diagram}. 
These masses are calculated as
\begin{align}
m_\ell &= -\frac{f_L^\ell f_R^\ell s_{2\theta}}{32\pi^2}M_\ell \left[F \left(x_1^2\right) - F\left(x_2^2\right)\right], \label{eq:ml_exact}  
\end{align}
where $x_i\equiv m_{\eta_i^\pm}/M_\ell$, and $F$ is a loop function: 
\begin{align}
F(x) = \frac{x}{1 - x}\ln x . 
\end{align}
Clearly, we see that a non-trivial mixing angle $\theta$ and a mass difference between $\eta_1^\pm$ and $\eta_2^\pm$ are required to obtain finite masses of charged leptons.  For $f_L^\ell f_R^\ell s_{2\theta} ={\cal O}(1)$ and $m_{\eta_{1,2}^\pm} = {\cal O}(100)$~GeV, the observed muon and electron masses are reproduced by taking $M_\mu$ and $M_e$ to be of order TeV and PeV, respectively.  This also means that the electron mass is reproduced by taking $f_L^e f_R^e s_{2\theta} = m_e/m_\mu \simeq 1/200$ for $M_\mu \approx M_e ={\cal O}(1)$~TeV, which can be considered as a natural choice, since the size of the couplings ($f_L^e$ and $f_R^e$) can be of order 0.1.  We note that the decoupling behavior of $m_\ell$ can be checked as $m_\ell \sim M_\ell\, v^2/M_{\rm max}^2$ for $M_{\rm max} \gg v$ with $M_{\rm max} \equiv \text{max}(M_\ell,m_{\eta_1^\pm},m_{\eta_2^\pm})$.

\begin{figure}[t]
\begin{center}
 \includegraphics[width=80mm]{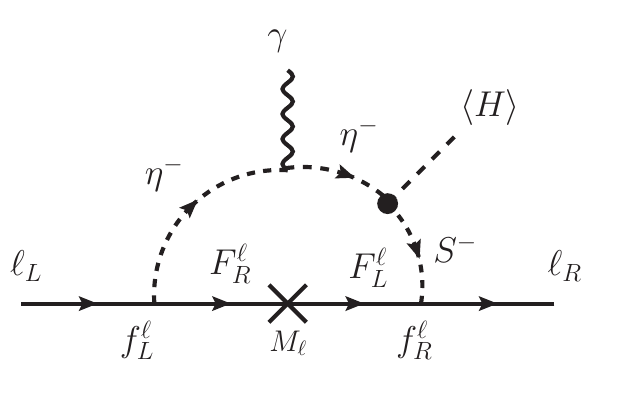}
   \caption{Dominant new contributions to $(g-2)$ for charged leptons. }
   \label{fig:diagram2}
\end{center}
\end{figure}

New contributions to electron and muon $(g-2)$'s are obtained from the diagram shown in Fig.~\ref{fig:diagram2}. 
There are the other contributions to the $(g-2)$'s without the Higgs VEV insertion into the scalar loop.  In such cases, the charged lepton chirality is flipped by the mass insertion at one of the external charged leptons, so that these contributions are negligibly smaller than that given in Fig.~\ref{fig:diagram2}.  It is clear that the structure of the couplings is the same as that of the mass generation for the corresponding charged leptons.  As such, the coupling dependence can be replaced in terms of the charged lepton mass as~\cite{Chiang:2021pma}: 
\begin{align}
\begin{split}
\Delta a_\ell &\equiv a_\ell^{\rm New} - a_\ell^{\rm SM} 
= \frac{2m_\ell^2}{M_{\ell}^2}\frac{G\left(x_1^2 \right) - G\left(x_2^2 \right)}{F\left(x_1^2 \right)- F\left(x_2^2 \right)}
~, 
\label{eq:da_app}
\end{split}
\end{align}
where 
\begin{align}
G(x) = \frac{1-x^2+2x \ln x }{ 2(1-x)^3}
~. \label{eq:g}
\end{align}
It is important to note here that the new contributions to muon and electron $(g-2)$'s are always positive.  This is in line with the current $\sim 4.2\sigma$ discrepancy between the SM prediction and the experimental result given by Fermilab~\cite{Abi2021}.  We adopt the following values for the differences between experimental values and the SM prediction for the electron $(g-2)$~\cite{Morel2020} and the muon $(g-2)$~\cite{Abi2021}: 
\begin{align}
\begin{split}
\Delta a_\mu^{\rm Exp} &\equiv a_\mu^{\rm Exp} - a_\mu^{\rm SM} = (251 \pm 59)\times 10^{-11}
~,
\\ 
\Delta a_e^{\rm Exp}  &\equiv a_e^{\rm Exp} - a_e^{\rm SM} = (4.8 \pm 3.0)\times 10^{-13}
~. 
\end{split}
\label{eq:dae}
\end{align}
We note that the above $\Delta a_e^{\rm Exp}$, which indicates consistency with the SM prediction, is obtained from the measurement of $\alpha_{\rm em}$ by using the rubidium atom~\cite{Morel2020}.  
It has been known that the measurement of $\alpha_{\rm em}$ by using the cesium atom~\cite{Jegerlehner2018} results in a significantly different value of $\Delta a_e^{\rm Exp}=(-8.7 \pm 3.6)\times 10^{-13}$ from that given in Eq.~(\ref{eq:dae}), but it is not considered in this paper \footnote{The negative discrepancy in the electron $g-2$ measured from cesium atoms cannot be explained in the present model because the signs of $\Delta a_e$ and $\Delta a_\mu$
are determined to be both positive. 
On the other hand, the negative (positive) discrepancy in the electron (muon) $g-2$ can simultaneously be explained in the model proposed in Ref.~\cite{Chen:2020tfr} 
(see also Ref.~\cite{Calibbi:2018rzv} for a similar model construction) whose model setup is quite similar 
to the present model, where the masses of charged leptons are given at tree level. }.  In Eq.~(\ref{eq:da_app}), we see that $M_\mu$ is almost determined by requiring $\Delta a_\mu$ to fall within, {\it e.g.}, the 2$\sigma$ interval given in Eq.~(\ref{eq:dae}).  In fact, $M_\mu$ should be between $2-3.5$~TeV in order to explain the muon $(g-2)$ anomaly at 2$\sigma$ level as shown in Ref.~\cite{Chiang:2021pma}.  
On the other hand, $\Delta a_e^{\rm Exp}$ is perfectly consistent with the SM prediction, so that we obtain only a lower limit on $M_e$: about 300~GeV for $m_{\eta_{1,2}^{\pm}} = {\cal O}(100)$~GeV.

In our model, Majorana mass for left-handed neutrinos is also generated at the one-loop level, where the right-handed neutrinos run in the loop as shown in the right plot of Fig.~\ref{fig:diagram}.   
This diagram is the same as that in the so-called scotogenic model proposed in Ref.~\cite{Ma:2006km}.  
The mass matrix for left-handed neutrinos is given by
\begin{align}
({\cal M}_\nu)_{\ell \ell'} = \frac{1}{32\pi^2}\sum_{i = 1,2,3}y_R^{\ell i}M_R^i y_R^{\ell' i}\left[F\left(\frac{m_{\eta_R}^2}{(M_R^i)^2}\right) - F\left(\frac{m_{\eta_I}^2}{(M_R^i)^2}\right)\right]
~. \label{eq:ne}
\end{align}
This expression is approximately expressed for $M_R^{i} = M_R$ and $\lambda_5 \ll 1$ as
\begin{align}
({\cal M}_\nu)_{\ell \ell'} &\simeq 
\begin{cases}
\displaystyle
-\frac{M_R}{32\pi^2}\frac{v^2\lambda_5}{m_{\eta^0}^2}\sum_{i = 1,2,3}y_R^{\ell i} y_R^{\ell' i}~~\text{for}~~M_R \ll m_{\eta^0} \\
\displaystyle
-\frac{1}{64\pi^2}\frac{v^2\lambda_5}{M_R}\sum_{i = 1,2,3}y_R^{\ell i} y_R^{\ell' i}~~\text{for}~~M_R \sim m_{\eta^0}
\end{cases},
\label{eq:ne1}
\end{align}
where $m_{\eta^0}^2 \equiv (m_{\eta_R}^2 + m_{\eta_I}^2)/2$.  This matrix can be diagonalized by introducing the Pontecorvo-Maki-Nakagawa-Sakata (PMNS) matrix $U_{\rm PMNS}$ as follows:
\begin{align}
U_{\rm PMNS}^T\, {\cal M}_\nu\,  U_{\rm PMNS} = \text{diag}(m_1,m_2,m_3)
~, \label{eq:ne2}
\end{align}
where $m_i$ ($i=1,2,3)$ are the mass eigenvalues of the left-handed neutrinos.  These masses are usually expressed in terms of the smallest eigenvalue $m_0 \equiv  m_1$ ($m_3$) and two squared mass differences $\Delta m_{\rm sol}^2 \equiv m_2^2 - m_1^2$ and $\Delta m_{\rm atm} \equiv m_3^2 - m_1^2$ ($m_2^2 - m_3^2$) assuming the normal (inverted) hierarchy for the neutrino mass spectrum.  Generally, $U_{\rm PMNS}$ is expressed in terms of three mixing angles $\theta_{12}$,  $\theta_{23}$, $\theta_{13}$ and three CP phases.  Thus, the neutrino mass matrix ${\cal M}_\nu$ is written in terms of $\{m_0,\Delta m_{\rm sol}^2,\Delta m_{\rm atm}^2, \theta_{12},\theta_{23},\theta_{13}\}$ via Eq.~(\ref{eq:ne2}), if we neglect the CP phases in the lepton sector Lagrangian.  It is clear from Eq.~(\ref{eq:ne}) that we can easily accommodate the observed data of neutrino oscillations by properly choosing the elements of $y_R^{}$, which is a general $3\times 3$ real matrix.  From Eq.~(\ref{eq:ne1}) and taking $y_R^{ij} = \delta^{ij}y_R^{}$, 
we can estimate the typical size of the neutrino mass as:
\begin{align}
m_\nu \simeq 
\begin{cases}
\displaystyle
0.1~\text{eV}\times \left(\frac{M_R}{100~\text{MeV}}\right) \times \left(\frac{200~\text{GeV}}{m_{\eta^0}}\right)^2\times\left(\frac{y_R^{}}{10^{-3}}\right)^2\times \frac{\lambda_5}{0.2}~~\text{for}~~M_R \ll m_{\eta^0} 
\\
\displaystyle
0.1~\text{eV}\times \left(\frac{200~\text{GeV}}{M_R}\right) \times\left(\frac{y_R^{}}{10^{-4}}\right)^2\times \frac{\lambda_5}{0.02}~~\text{for}~~M_R \sim m_{\eta^0}
\end{cases},
\label{eq:nu10}
\end{align}
with $m_i = m_\nu$.

\begin{figure}[t]
\begin{center}
 \includegraphics[width=80mm]{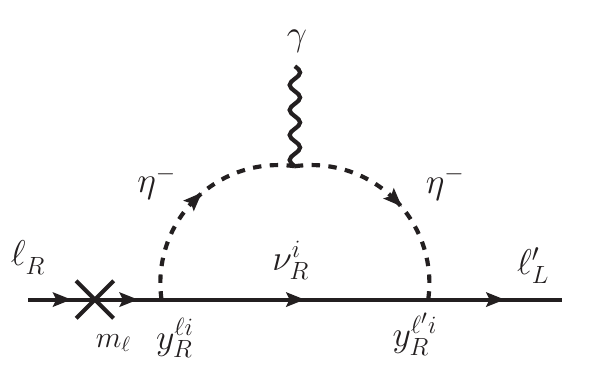}
   \caption{Feynman diagrams for LFV decays of charged leptons with $\ell \neq \ell'$. }
   \label{fig:diagram3}
\end{center}
\end{figure}

The coupling $y_R^{}$ generally leads to radiative LFV decays of charged leptons~\cite{Toma:2013zsa,Vicente:2014wga,Kubo:2006yx,Ma:2001mr}, as shown in Fig.~\ref{fig:diagram3}.  
The branching ratios for $\ell \to \ell' \gamma$ ($\ell \neq \ell'$) decays are calculated as 
\begin{align}
{\cal B}(\ell \to \ell' \gamma) & = \frac{3\alpha_{\rm em}}{64\pi G_F^2}\left|\sum_{i=1,2,3}y_R^{\ell i} y_R^{\ell' i}\left[ 
\frac{c_\theta^2}{m_{\eta_1^\pm}^2}H\left(\frac{(M_R^i)^2}{m_{\eta_1^\pm}^2}\right) +\frac{s_\theta^2}{m_{\eta_2^\pm}^2}H\left(\frac{(M_R^i)^2}{m_{\eta_2^\pm}^2}\right) 
\right] \right|^2C_{\ell\ell'}
~, \label{eq:lfv}
\end{align}
where 
\begin{align}
C_{\tau\ell} = \tau_{\tau}\times \Gamma(\tau \to \ell \bar{\nu}\nu) \simeq 0.174 ~ (0.179)~\text{for}~\ell = \mu (e),~~\text{and}~~C_{\mu e} = 1
~, 
\end{align}
with $\tau_\tau$ and $\Gamma(\tau \to \ell \bar{\nu}\nu)$ being the lifetime and the leptonic partial decay rate of the tau lepton, respectively.  The loop function $H(x)$ is given by 
\begin{align}
H(x)& =  \frac{1-6x+3x^2+2x^3-6x^2\ln x}{6(1-x)^4}
~. 
\end{align}
We confirm that the above expression is consistent with that given in Ref.~\cite{Ma:2001mr} for $\theta = 0$.  
When $M_R^i \ll m_{\eta_1^\pm},m_{\eta_2^\pm}$, the branching ratio takes a simple form as 
\begin{align}
{\cal B}(\mu \to e \gamma) &\simeq
\frac{\alpha_{\rm em}}{768\pi G_F^2}(\bar{y}_R^{\mu e})^4\left(\frac{c_\theta^2}{m_{\eta_1^\pm}^2} +\frac{s_\theta^2}{m_{\eta_2^\pm}^2}\right)^2 
\simeq 1.4\times 10^{-13}\times \left(\frac{\bar{y}_R^{\mu e}}{0.01}\right)^4 \times \left(\frac{200~\text{GeV}}{m_{\eta^\pm}}\right)^4
~, 
 \label{eq:al3}
\end{align}
where $\bar{y}_R^{\mu e} \equiv \sqrt{ |\sum_{i=1,2,3}y_R^{\mu i}y_R^{e i}|}$ and $m_{\eta^\pm} = m_{\eta_1^\pm} = m_{\eta_2^\pm}$ in the right-most expression.  For the case with $M_R \sim m_{\eta^\pm}$, a factor of 1/4 is further multiplied to the above expression.  Therefore, the current upper limit ${\cal B}(\mu \to e\gamma) < 4.2 \times 10^{-13}$ at 90\% confidence level (CL) from the MEG experiment~\cite{MEG:2016leq} can easily be avoided by taking the typical values of the parameters to reproduce the neutrino mass in Eq.~(\ref{eq:nu10}). 

We note in passing that there are generally $\ell \to \ell'\ell''\ell'''$ type LFV decays of charged leptons such as $\mu \to 3e$.  
In our model, these processes occur 
via penguin type and box type diagrams with $\nu_R^{}$ and $\eta_{1,2}^\pm$ running in the loops, and their contributions to the decay rate are proportional to $\alpha_{\rm em}^2\,y_R^4$ and $y_R^8$, respectively. 
Thus, for $y_R \sim {\cal O}(10^{-3})$ or smaller, the branching ratios of these processes are negligibly small as compared 
with the current bound, {\it e.g.}, ${\cal B}(\mu \to 3e) < 1.0\times 10^{-12}$ at 90\% CL~\cite{SINDRUM:1987nra}.

\section{Collider Phenomenology \label{sec:collider}}

In this section, we discuss constraints on the parameter space from current LHC data and the prospect of testing the model in future collider experiments.

\subsection{Constraint from direct searches at LHC}

Our model has several new particles in the dark sector, {\it i.e.}, the vector-like leptons $F^\ell$, the right-handed neutrinos $\nu_R^i$, and the dark scalars $\eta_{R,I}^{}$, $\eta_{1,2}^\pm$.  As discussed in Sec.~\ref{sec:lepton}, the masses of $F^\ell$ are typically of order TeV or higher, and they can only interact with electrons or muons via the new Yukawa couplings $f_{e,\mu}^{\ell}$.  As such, it is quite challenging to directly search for $F^{\ell}$ at the LHC.  Therefore, we focus on the phenomenology of the dark scalar bosons and $\nu_R^{i}$.  In the following discussions, we take $m_{\eta_2^\pm} \geq m_{\eta_1^\pm} \geq  m_{\eta_R}$, and $m_{\eta_I}$ is fixed such that the $\Delta T$ parameter, given in Eq.~(\ref{eq:deltat}), is identically zero, which can be satisfied by taking $m_{\eta_2^\pm}  \geq m_{\eta_I} \geq m_{\eta_1^\pm}$.  In addition, we assume degenerate masses for the right-handed neutrinos $M_R^i = M_R$.

At the LHC, the dark scalars are produced in pair via the $s$-channel gauge boson exchange, {\it i.e.}, $pp \to \gamma^*/Z^* \to \eta_{1,2}^\pm \eta_{1,2}^\mp$ and $pp \to W^{\pm *} \to \eta_{1,2}^\pm \eta_{R,I}^{}$. 
Their decay modes can be classified into three types: (i) decays via gauge couplings, {\it i.e.},
\begin{align}
\eta_2^\pm \to W^{\pm (*)}\eta_{R,I}^{}
~,\quad 
\eta_2^\pm \to Z^{(*)}\eta_{1}^\pm
~,\quad 
\eta_I^{} \to  W^{(*)}\eta_{1}^{\pm}
~,\quad 
\eta_I^{} \to  Z^{(*)}\eta_{R}^{}
~, \quad
\eta_1^\pm \to W^{\pm (*)}\eta_{R}^{}
~;  \label{eq:gauge-decay}
\end{align}
(ii) those via the Yukawa coupling $y_R$, {\it i.e.},
\begin{align}
\eta_{1,2}^\pm \to \ell_L^{\pm} \nu_R^{}
~, \quad 
\eta_{R,I}^{} \to \nu_L^{} \nu_R^{}
~, \label{eq:yukawa-decay}
\end{align}
and (iii) those via the scalar trilinear coupling $\lambda_{h\eta_1^\pm \eta_2^\mp}$, see Eq.~(\ref{eq:lam12h}), {\it i.e.},
\begin{align}
\eta_2^\pm \to \eta_1^\pm h
~, \label{eq:higgs-decay}
\end{align}
where the modes in (ii) are kinematically possible if $M_R$ is smaller than the mass of the dark scalars.  The former decay modes are severely constrained by an appropriate reinterpretation of the searches for chargino-neutralino pair productions at the LHC with the integrated luminosity of 139 fb$^{-1}$~\cite{ATLAS:2021moa}, where these SUSY particles are assumed to decay into the lightest neutralino (dark matter) and a weak boson or a Higgs boson.  
For the case where the chargino and the neutralino mainly decay into a weak boson and dark matter, 
the lower limit on their masses is found to be about 650~GeV at 95\% CL~\cite{ATLAS:2021moa} when the dark matter mass is of order 100~GeV.  
This strong bound can be applied to our model as explained below. 

On the other hand, the decays induced by the Yukawa coupling given in Eq.~(\ref{eq:yukawa-decay}) can also be constrained by the slepton searches at the LHC~\cite{ATLAS:2019lff,ATLAS:2019gti}. \footnote{See Refs.~\cite{Cao:2017ffm,Babu:2019mfe} for the detectability of singly-charged scalar bosons mainly decaying into a charged lepton and a neutrino in future collider experiments.  } 
From the dataset with the integrated luminosity of 139 fb$^{-1}$, the mass of sleptons $m_{\tilde{\ell}}$ has been constrained to be $100 \lesssim m_{\tilde{\ell}} \lesssim 600$~GeV ($120 \lesssim  m_{\tilde{\ell}} \lesssim 390$~GeV) for $\tilde{\ell} = \tilde{e}$ or $\tilde{\mu}$ ($\tilde{\tau}$) as long as the lightest neutralino and sleptons are not (nearly) degenerate in mass.\footnote{
According to Refs.~\cite{ATLAS:2019lff,ATLAS:2019gti}, we can extract an upper bound on the mass difference between a slepton and a neutralino to be about 50 (100) GeV from the $\tilde{e}$ or $\tilde{\mu}$ ($\tilde{\tau}$) search.  
}. 
Similar events are expected in our model as 
\begin{align}
pp \to \gamma^*/Z^* \to \eta_{1,2}^{\pm}\eta_{1,2}^{\mp} \to \ell\ell'\nu_R^{}\nu_R^{}
~. \label{eq:slepton}
\end{align}

The phenomenology can be drastically changed, depending on the mass spectrum of the dark sector particles.  In the following, we discuss two scenarios referred to as Scenario-I and Scenario-II, in which $\nu_R$ are assumed to be heavier and lighter than the dark scalar bosons, respectively. 

Let us first consider Scenario-I, where $\eta_R^{}$ can be a dark matter candidate.  As shown in Ref.~\cite{Chen:2020tfr}, the relic abundance of dark matter can be explained by taking $m_{\eta_R} \simeq 63$~GeV or $m_{\eta_R} \gtrsim 80$~GeV.  The dark matter $\eta_R^{}$ can interact with nucleus via the Higgs boson exchange, and its amplitude is proportional to the $\eta_R^{}\eta_R^{}h$ coupling.  We thus define the dark matter-Higgs coupling $\lambda_{\rm DM}$ by ${\cal L} \ni v\lambda_{\rm DM} \eta_R^{}\eta_R^{}h$, which is extracted to be 
\begin{align}
\lambda_{\rm DM}= \frac{m^2_{\eta_1^\pm}}{v^2}c_\theta^2 + \frac{m^2_{\eta_2^\pm}}{v^2}s_\theta^2 - \frac{m^2_{\eta_R}}{v^2} - \frac{\lambda_3}{2}
~. 
\end{align}
It has been shown in Ref.~\cite{Chen:2020tfr} that $\lambda_{\rm DM}\gtrsim 3\times 10^{-3}$ has been excluded  from the XENON1T experiment~\cite{Aprile:2018dbl}.  For concreteness, we take $m_{\eta_R}=63$~GeV and  $\lambda_{\rm DM} =10^{-3}$ in Scenario-I.

With regard to the constraint from LHC data, we take into account the bound from the neutralino-chargino pair production explained above.  
In Ref.~\cite{ATLAS:2021moa}, upper limits on the cross section are given depending on several signal regions.
Here, we employ the upper limit of $0.04$ fb given for the signal region {\tt incSR$^{\tt WZ}$-2} with on-shell $W$ and $Z$ bosons defined in Ref.~\cite{ATLAS:2021moa}, and apply it to our model in the following way:
\begin{align}
\sum_{i=1,2}\sigma(pp \to \eta_i^\pm \eta_I^{} \to WZ\eta_R^{}\eta_R^{}) \times \epsilon \leq  0.04~\text{fb}, \label{eq:bound01}
\end{align}
where $\epsilon$ is an efficiency factor whose value is determined by the distributions of various kinematical observables such as missing transverse energies $E_T^{\rm miss}$, the invariant mass for a dilepton system $m_{\ell\ell}$, the number of jets, and so on. 
We note that the cross section of (wino-like) neutralino-chargino pair productions for their degenerate mass of 650~GeV, corresponding to the lower limit taken in Ref.~\cite{ATLAS:2021moa},  
at the 13-TeV LHC is given to be about 14~fb at next-to-leading order with next-to-leading logarithm in QCD~\cite{Fiaschi:2018hgm}. 
Thus, we obtain $\epsilon \simeq 0.3\%$ by naive estimation. 
In the following discussion, we take $\epsilon$ to be 0.3\%, 3\%, 30\% and 100\%, where the larger values of $\epsilon$ are to see stronger constraints expected to be obtained in future updates of data. 


\begin{figure}[t]
\begin{center}
 \includegraphics[width=80mm]{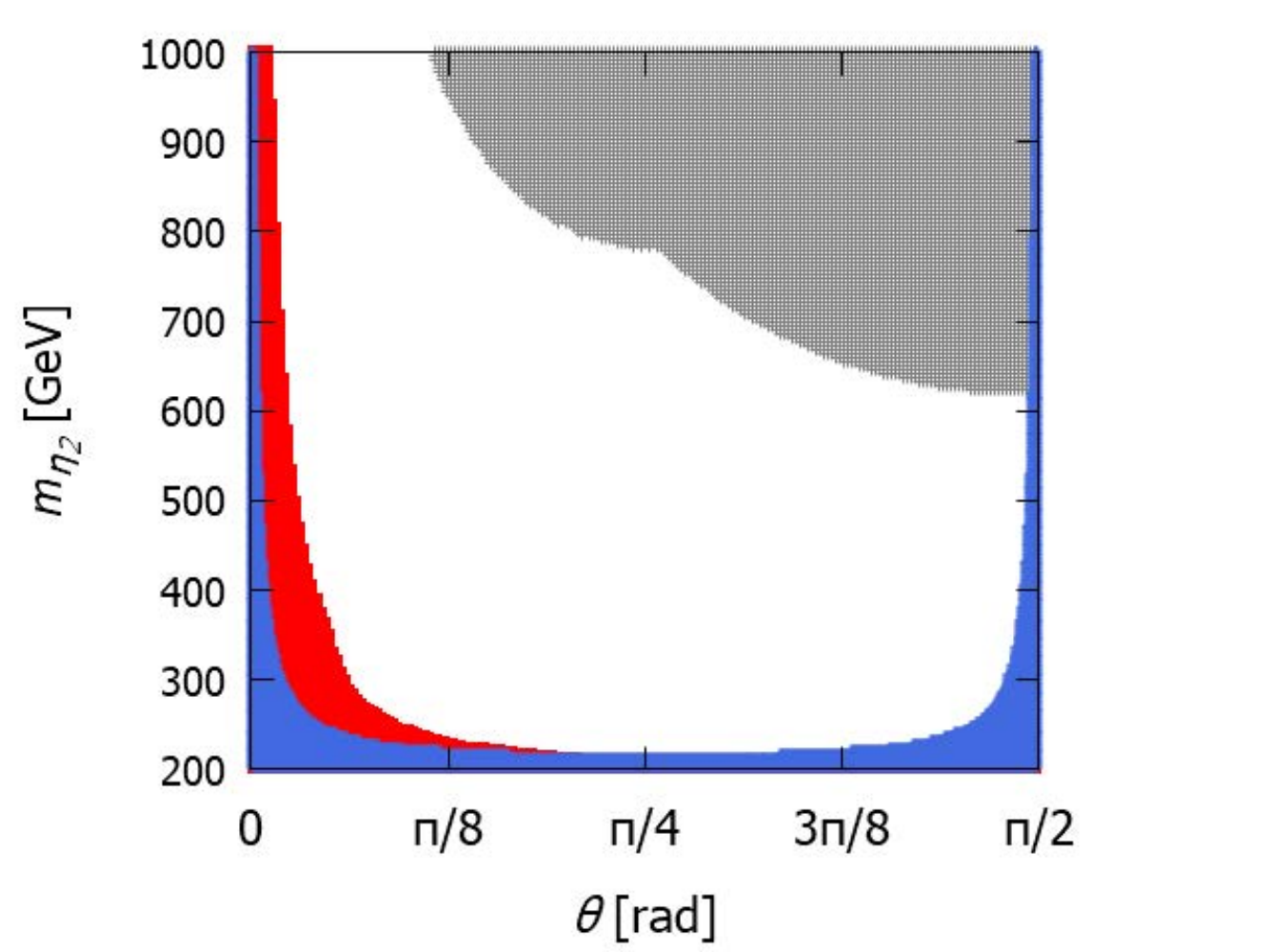}
 \includegraphics[width=80mm]{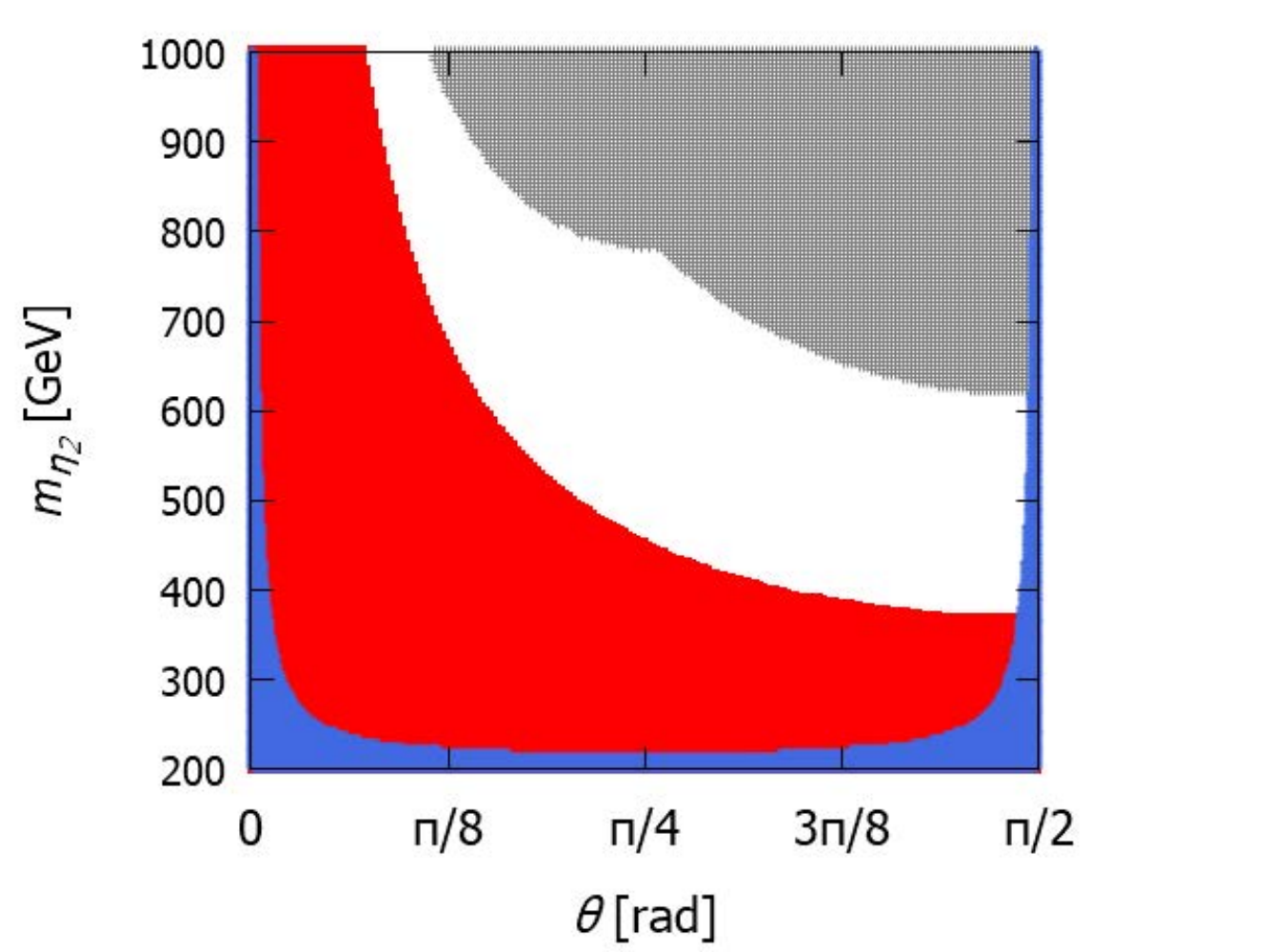}
 \includegraphics[width=80mm]{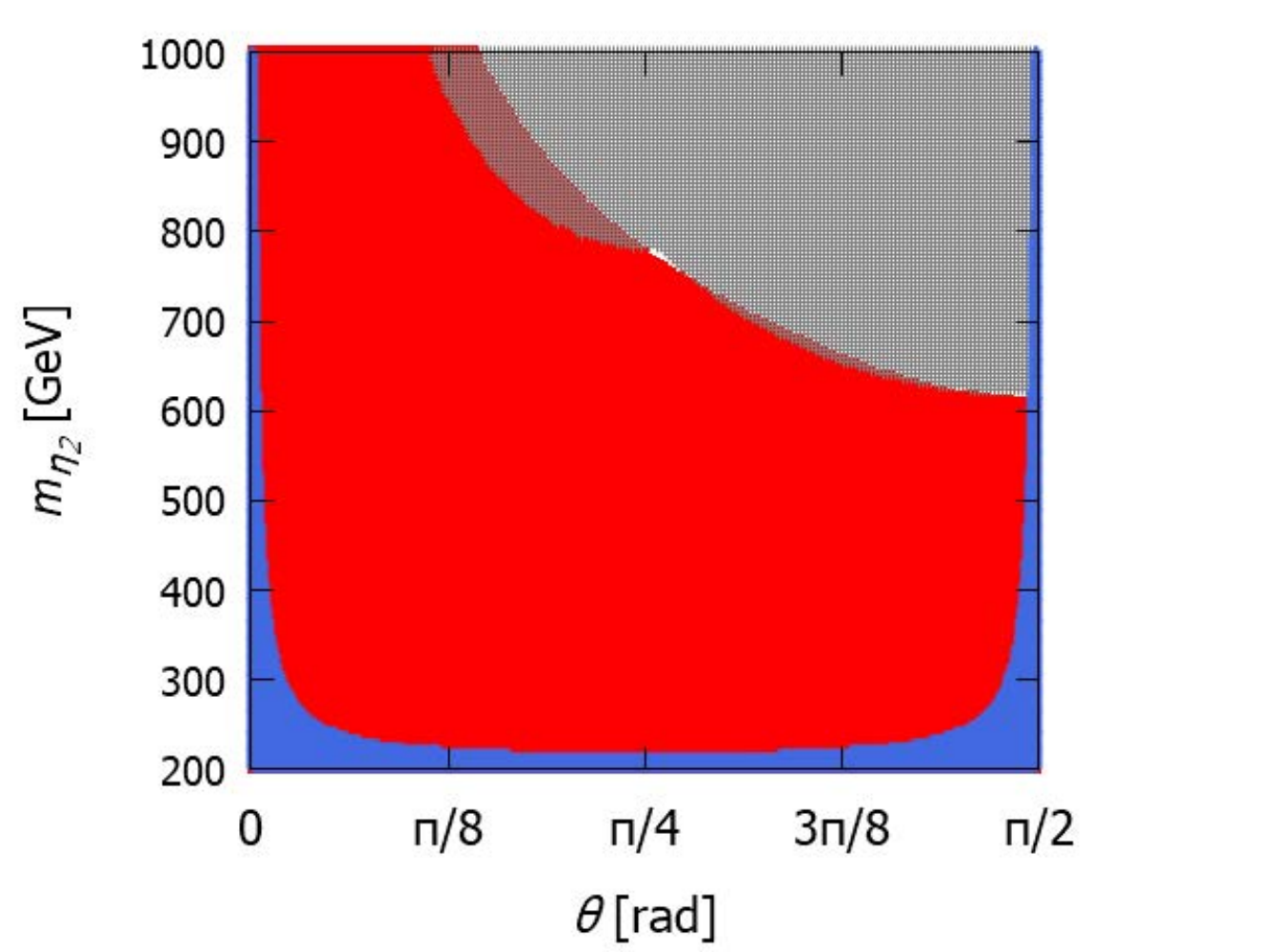}
 \includegraphics[width=80mm]{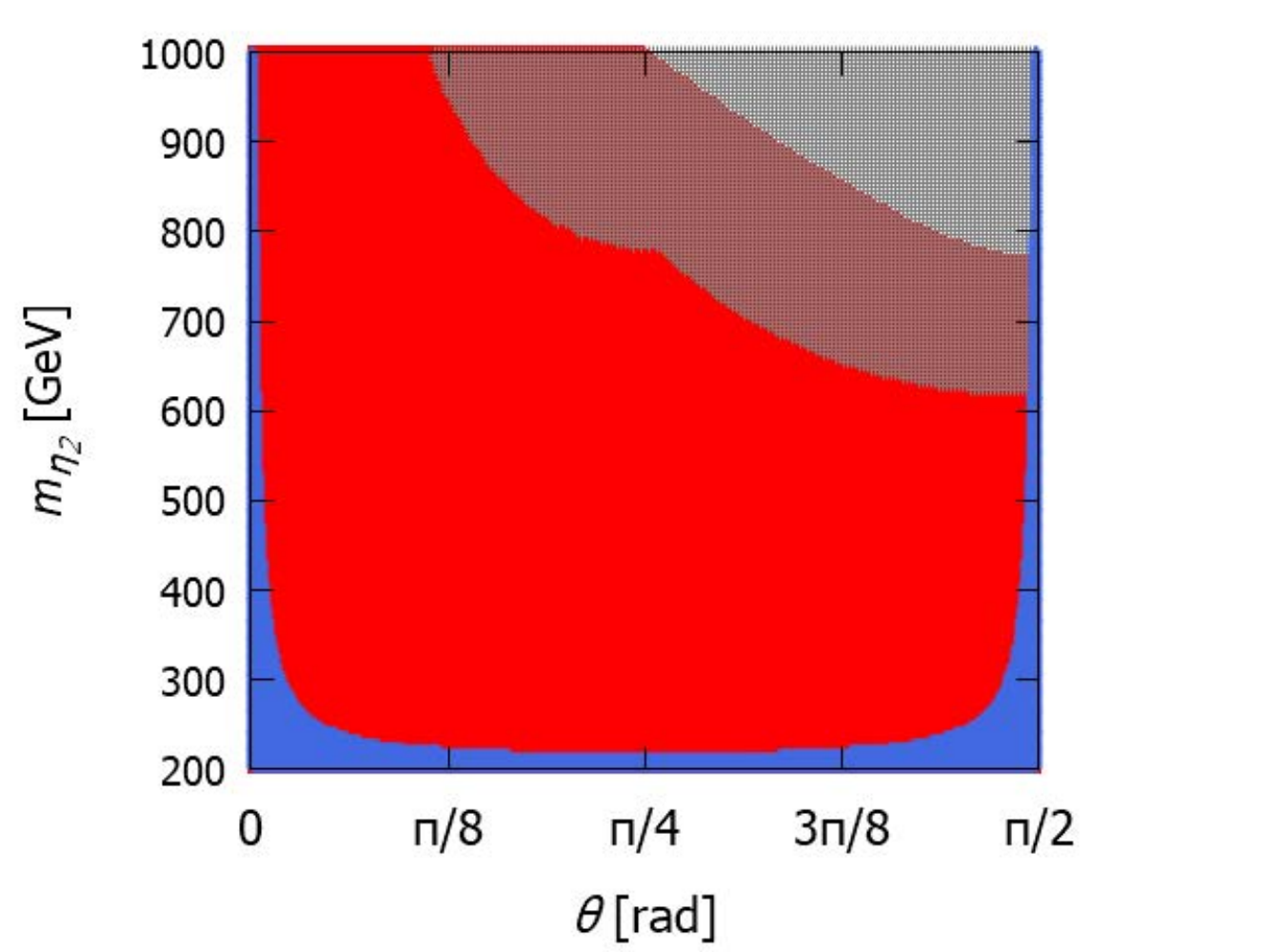}
\caption{Excluded regions in the $\theta$--$m_{\eta_2^\pm}$ plane for $m_{\eta_R}= 63$~GeV, $m_{\eta_1^\pm} =200$~GeV and $\lambda_{\rm DM} = 10^{-3}$.  
The efficiency factor $\epsilon$ is taken to be 0.3\% (value obtained by naive estimation: upper-left), 3\% (upper-right), 30\% (lower-left) and 100\% (lower-right). 
The value of $m_{\eta_I}$ is fixed so as to ensure $\Delta T = 0$.  
The other input parameters $\lambda_{2,6,7,8}$ are taken to be zero in this plot.  
The regions shaded in gray, red and blue are excluded by the perturbative unitarity bound, the LHC data and the perturbativity bound $(f_L^{\mu}f_R^{\mu} \leq 4\pi)$ for $M_\mu = 2$ TeV, respectively. }
   \label{fig:lhc1}
\end{center}
\end{figure}

Fig.~\ref{fig:lhc1} shows the parameter region excluded by various constraints in Scenario-I.  
It is seen that the bound from the direct search at LHC (shaded in red) gets stronger for smaller values of $\theta$.\footnote{For $\theta = \pi/2$ and $m_{\eta_2^\pm} = 800$ GeV, the mass of $\eta_I^{}$ is determined to be 800 GeV from 
the condition of $\Delta T = 0$. In this case, the production cross section of $q\bar{q}' \to W^{\pm *}\to \eta_{2}^\pm \eta_I^{}~(\eta_{1}^\pm \eta_I^{})$, which is proportional to $s_\theta^2$ ($c_\theta^2$), is calculated to be about 0.031 fb (0), while $\eta_2^\pm$ and $\eta_I$ decay into $W^\pm \eta_R$ and $Z \eta_R$ at almost 100\%.  Thus, this case provides a slightly smaller value of the upper limit on the cross section with $\epsilon = 100\%$. }
This is because the lighter charged scalars $\eta_1^\pm$ become more doublet-like states ($\eta^\pm$), and so the production cross section tends to be larger.  
We also impose the bound from the perturbative unitarity as discussed in Sec.~\ref{sec:model}, by which the region shaded in gray is excluded.  
Unlike the bound from the direct search, it gives an upper limit on $m_{\eta_2^\pm}$, because larger mass differences among the dark scalars require some of $\lambda_i$ parameters in the potential to be larger. 
It is seen that the unitarity bound tends to be stronger when $\theta$ approaches to $\pi/2$ at which the heavier charged scalars $\eta_2^\pm$ become more doublet-like states $\eta^\pm$, 
so that the large mass differences among the doublet states are required in such a region.  
The region shaded in blue is excluded by the perturbativity bound, {\it i.e.}, $f_L^\ell f_R^\ell < 4\pi$, whose value is determined by the charged lepton mass given in Eq.~(\ref{eq:ml_exact}) for fixed values of $\theta$, $m_{\eta_{1,2}^\pm}$ and $M_\ell$.  We here take $M_\mu = 2$~TeV which is required to explain the $(g-2)_\mu$ anomaly, and impose the perturbativity bound for $\ell = \mu$.  
Clearly, if $\theta \simeq n\pi/2$ ($n$ being an integer), the product of the Yukawa couplings $f_L^\mu f_R^\mu$ has to be larger to attain the observed charged lepton masses. 
Combining all the constraints with $\epsilon = 0.3\%$ (obtained by naive estimation), 
regions with $\theta \neq 0$ are typically allowed, while for $\epsilon > 30\%$ almost all the regions in Scenario-I are expected to be excluded.

Next, let us examine Scenario-II with $M_R < m_{\eta_R}$, where the right-handed neutrino $\nu_R^{}$~\footnote{We can consider a small mass difference among $\nu_R^i$ such that it does not affect the flavor and collider phenomenologies.  We can then identify the lightest right-handed neutrino as a dark matter candidate, and denote it as $\nu_R^{}$. } can be a dark matter candidate.  
In this case, $\nu_R^{}$ can annihilate into SM leptons via $t$-channel diagrams with the dark scalar exchange.  
As shown in Ref.~\cite{Kubo:2006yx}, the typical size of the coupling constant $y_R$ is required to be of order one for the case with the masses of $\nu_R$ and the dark scalars to be of order 100 GeV in order to 
reproduce the observed relic abundance of dark matter, {\it i.e.}, $\Omega_{\rm DM} h^2 \simeq 0.12$~\cite{Aghanim:2018eyx}.
For $M_R \ll m_\eta$ with $m_\eta$ being the typical mass of dark scalar bosons, 
larger values of $y_R$ are needed, because the annihilation cross section is suppressed by the factor of $M_R^2/m_\eta^2$~\cite{Kubo:2006yx}. 
Therefore, the relic abundance of $\nu_R$ is typically much larger than the observed value, because the size of $y_R$ is typically of order $10^{-3}$ or smaller as we have seen in Sec.~\ref{sec:lepton}. 

One simple solution to avoid such an over abundance problem would be to introduce complex scalar singlets\footnote{In order to avoid the over abundance problem only, introduction of one $\varphi$ would be enough. 
On the other hand, more than one $\varphi$ would be required to accommodate the observed neutrino oscillations. } $\varphi$ with non-trivial $U(1)_\ell$ charges, by which the Majorana masses for $\nu_R^i$ are effectively given by the VEV of $\varphi$ through new Yukawa interactions $\varphi \overline{\nu_R^{ic}} \nu_R^i$, instead of 
the mass term $\overline{\nu_R^{ic}} \nu_R^i$.  In this case, we obtain new trilinear couplings, {\it i.e.}, $\overline{\nu_R^{ic}} \nu_R^i \varphi_{R,I}^{}$, where $\varphi_{R}^{}$ is the real component of $\varphi$ and $\varphi_{I}^{}$ is the imaginary component which can be identified with a pseudo-NG boson associated with a spontaneous $U(1)_{\ell}$ breaking.  Through these interactions, there appear additional $s$-channel annihilation processes for $\nu_R^{}$ such as $\nu_R \overline{\nu_R} \to \varphi_R \to \varphi_I^{}\varphi_I^{}$.  If the mass of $\varphi_R$ is taken to be around $2M_R$, the relic abundance can be explained due to the resonance effect.  In this paper, we do not actually introduce $\varphi$ for simplicity. 

In Scenario-II, the constraint from the direct search at the LHC can be weaker than that in Scenario-I, 
because the decay branching ratios in Eq.~(\ref{eq:gauge-decay}) are diluted by the other modes in Eq.~(\ref{eq:yukawa-decay}).  As mentioned above, the mass of $\eta_R$ is no longer constrained by the dark matter abundance.  We thus simply fix $m_{\eta_R} = m_{\eta_1^\pm}$.  For the case with the maximal mixing $\theta = \pi/4$, the mass of $\eta_I^{}$ is determined to be that of $\eta_2^\pm$ from the condition of $\Delta T = 0$.  In this setup, we impose 
\begin{align}\sigma(pp \to \eta_2^\pm \eta_I^{}/\eta_{2}^+\eta_2^- \to W^{(*)}Z^{(*)}\eta_L\eta_L)\times \epsilon \leq 
\begin{cases}
 0.04~\text{fb}~~(\text{for the on-shell $WZ$ case})\\
 0.03~\text{fb}~~(\text{for the off-shell $WZ$ case})
\end{cases}, \label{eq:bound2}
\end{align}
where the left-hand side of the above expression represents the sum of the cross section times branching ratios providing the $ W^{(*)}Z^{(*)}\eta_L\eta_L$ events with $\eta_L$ being $\eta_1^\pm$ or $\eta_R^{}$. \footnote{In Scenario-II, $\eta_1^\pm$ can decay into $\ell^\pm \nu_R$, so that the signature eventually differs from the neutralino-chargino search mentioned above. We here do not take into account such a difference, and simply add the cross section with $\eta_L = \eta_1^\pm$ in Eq.~(\ref{eq:bound2}). }  
We note that the pair production of $\eta_2^\pm$ can also contribute to the above process in addition to the associated production $\eta_2^\pm \eta_I^{}$. 
Similar to the discussion in Scenario-I, we impose the upper limit on the cross section to be 0.04~fb for the on-shell $WZ$ production.  For the case with off-shell $WZ$ productions, 17 signal regions have been taken into account in Ref.~\cite{ATLAS:2021moa}, and for each region the upper limit on the cross section has been given.  We here simply impose the strongest one, 0.03~fb, to the cross section. 
We deal with the efficiency factor $\epsilon$ in the same way as for Scenario-I. 
 
\begin{figure}[t]
 \hspace{-5mm}
 \includegraphics[width=53mm]{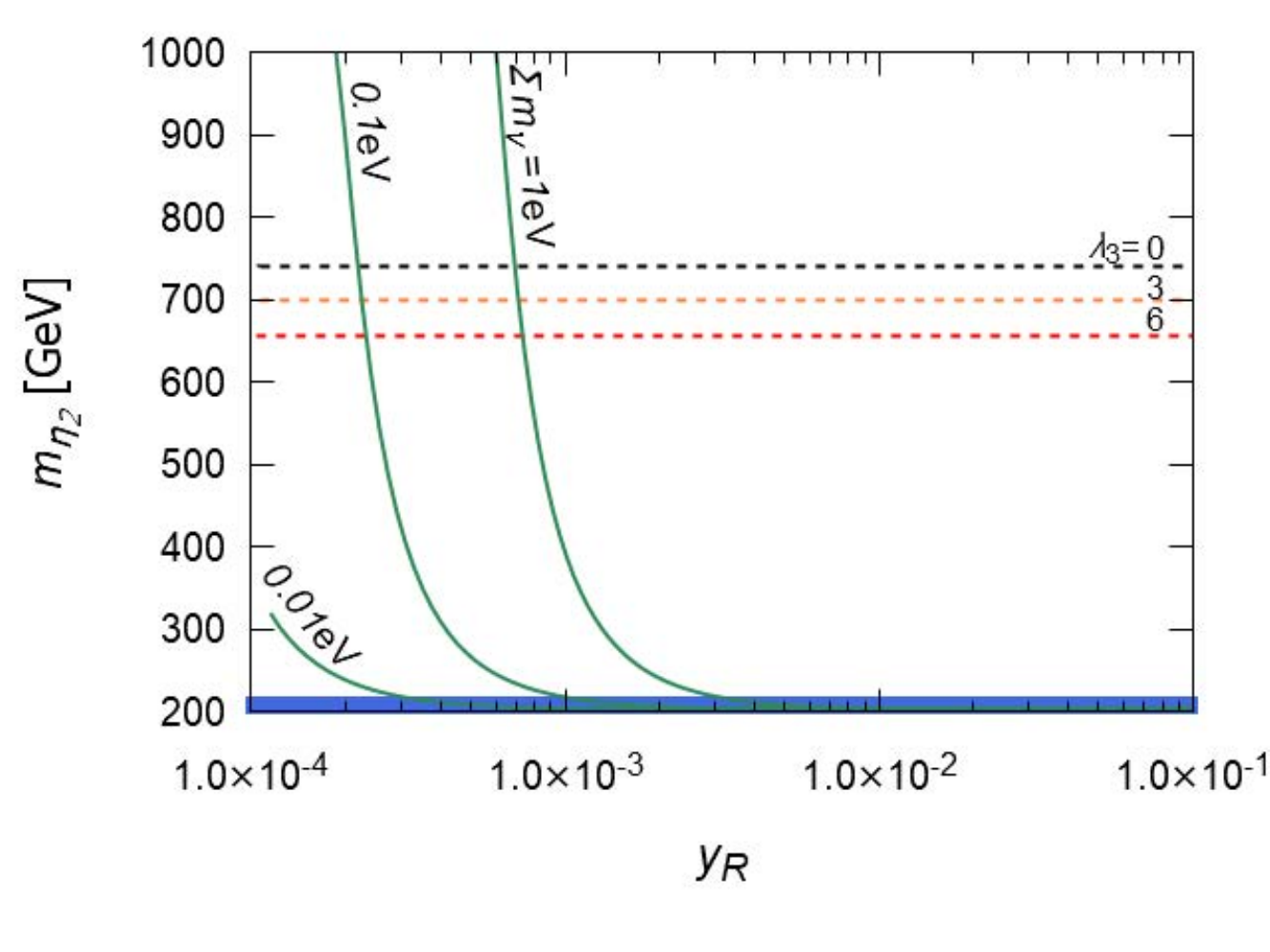}\hspace{-1mm}
 \includegraphics[width=53mm]{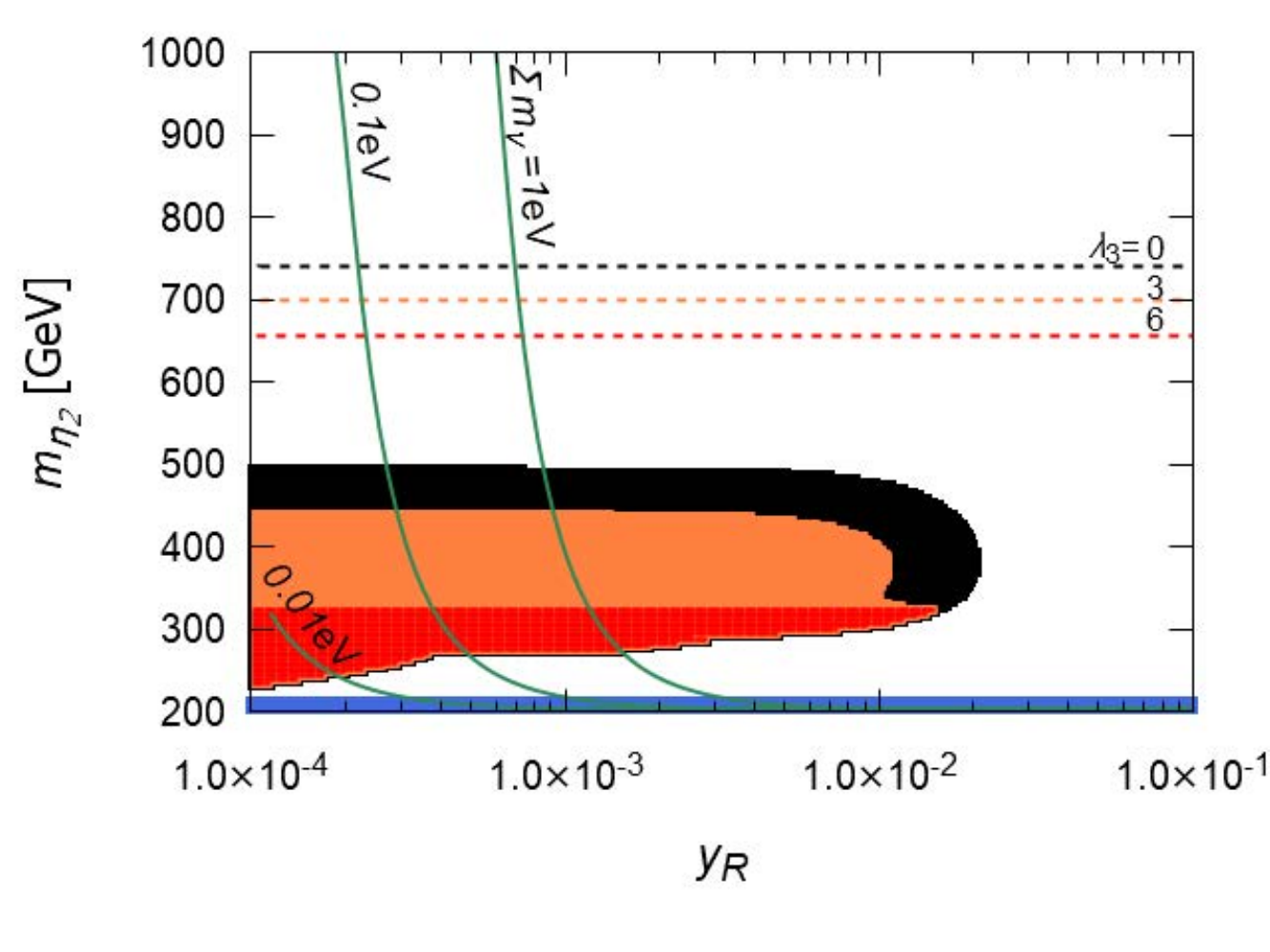}\hspace{-1mm}
 \includegraphics[width=53mm]{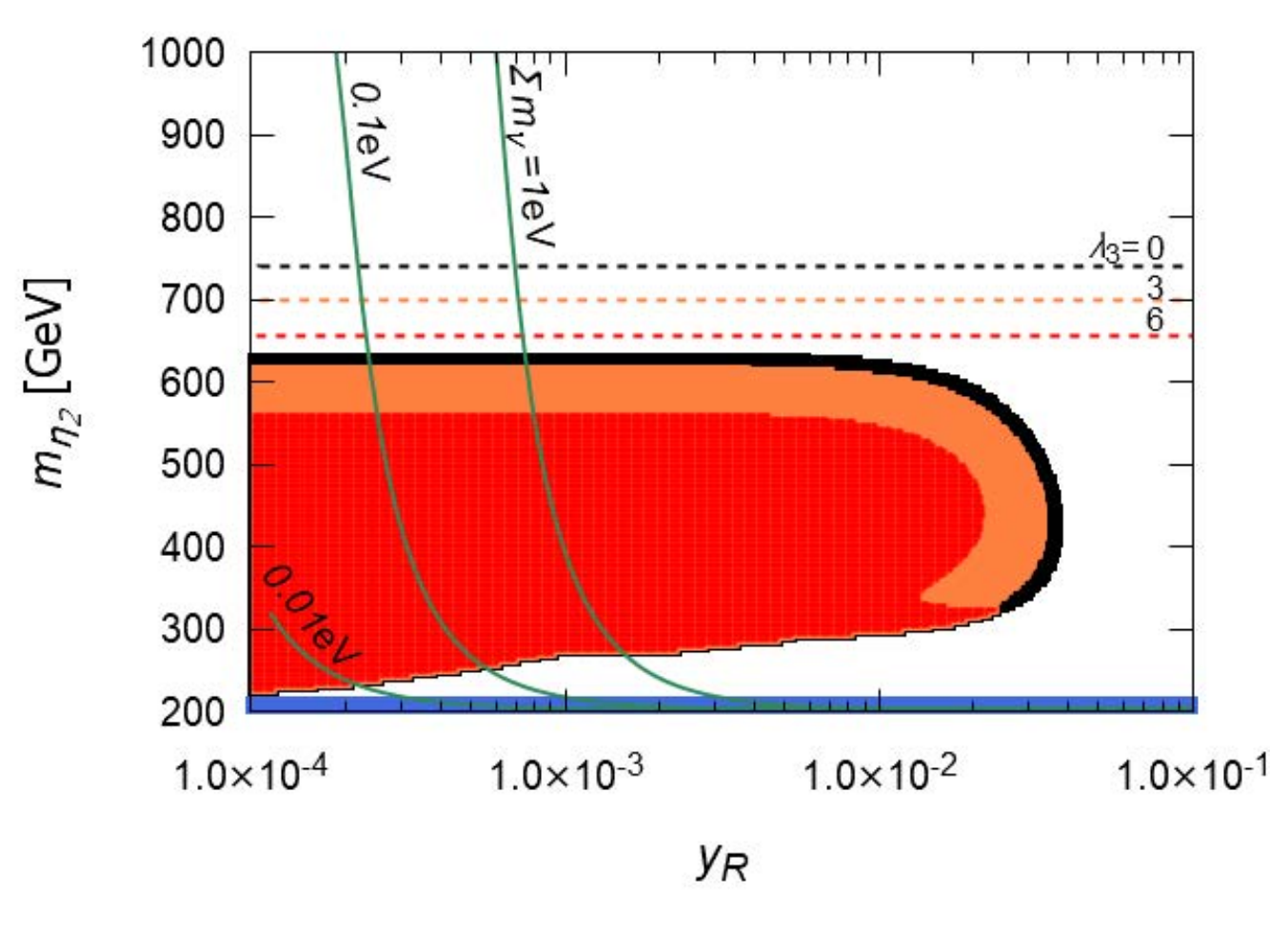} \\
 \hspace{-5mm}
 \includegraphics[width=53mm]{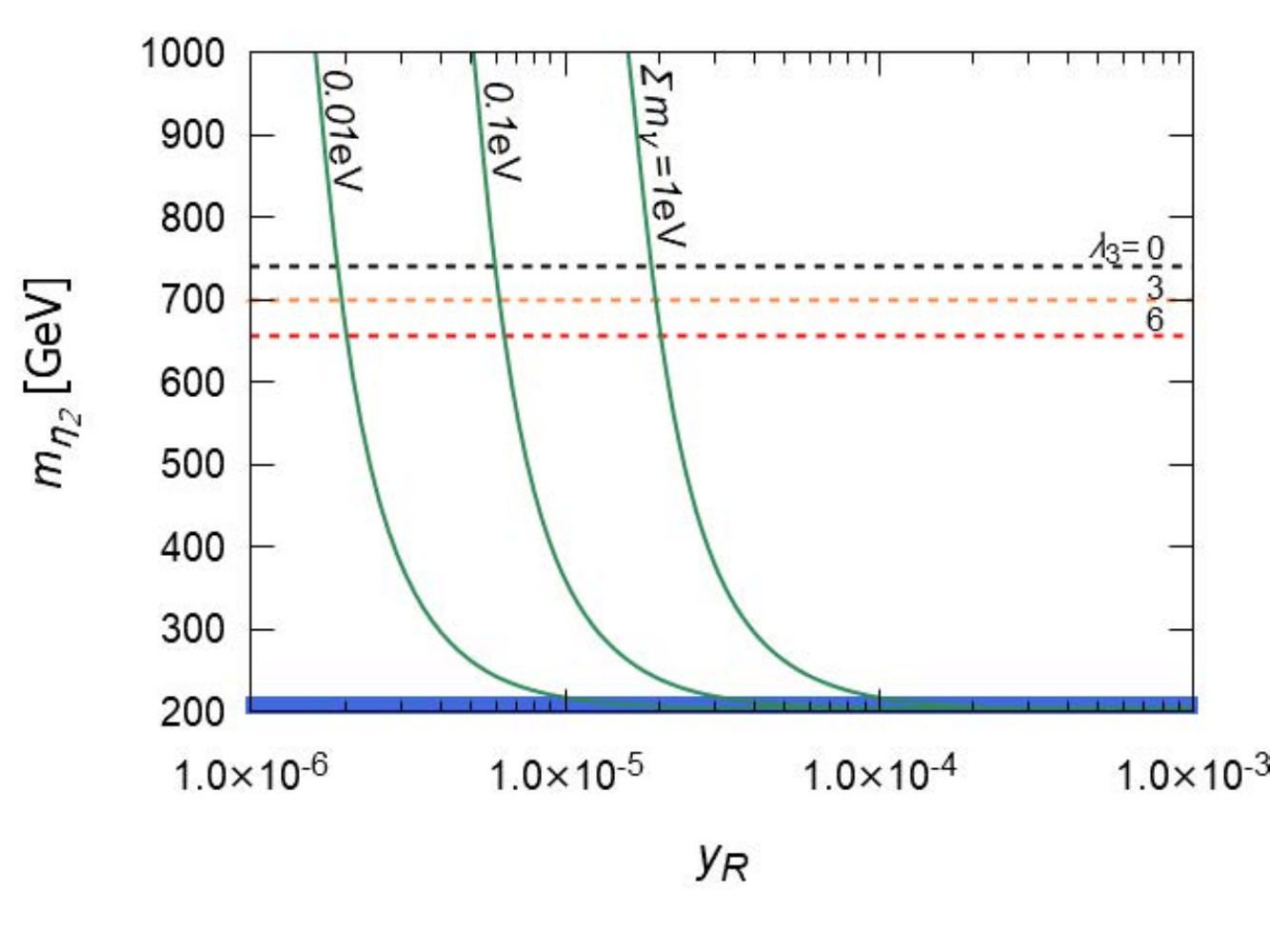}\hspace{-1mm}
 \includegraphics[width=53mm]{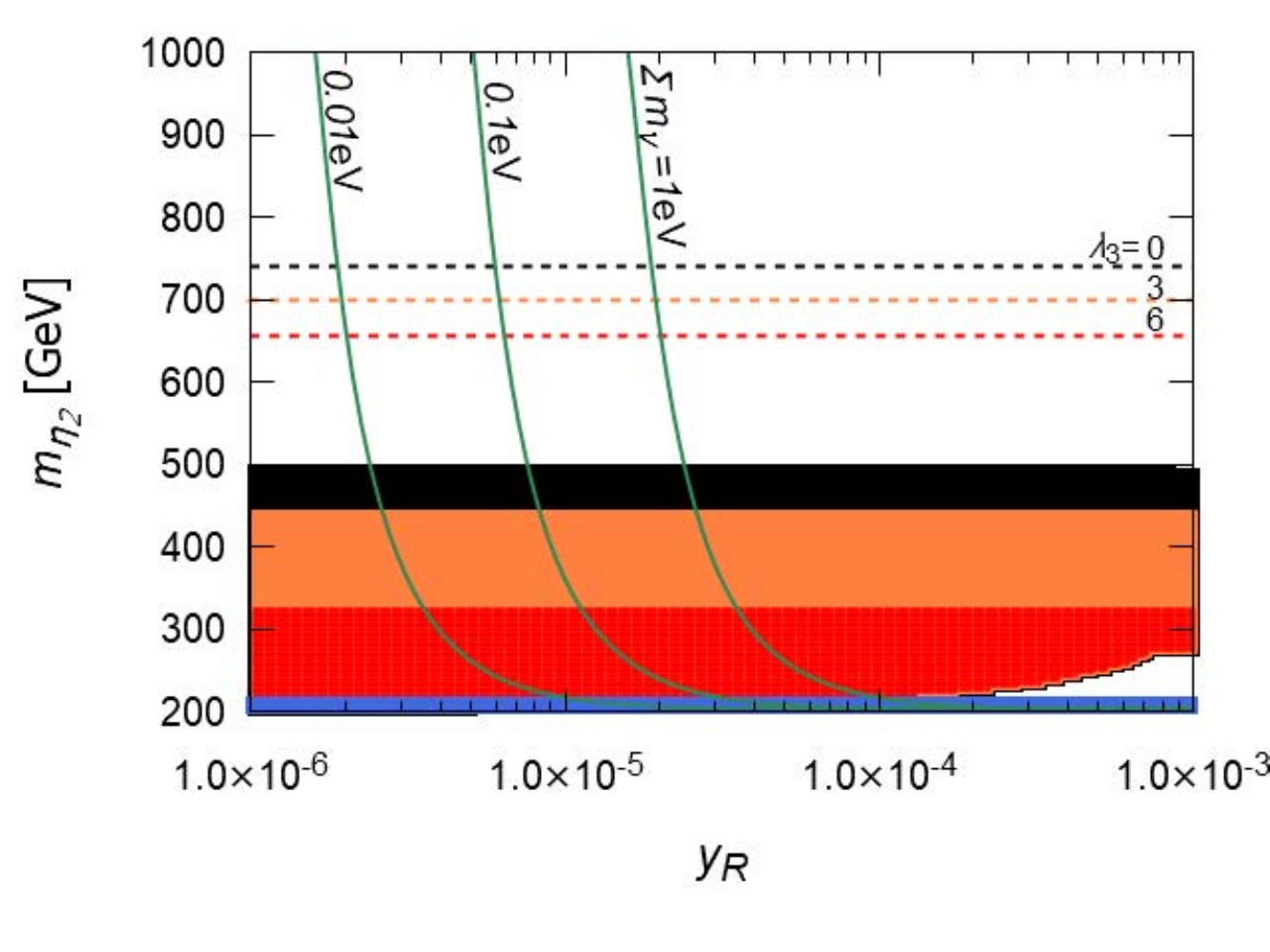}\hspace{-1mm}
 \includegraphics[width=53mm]{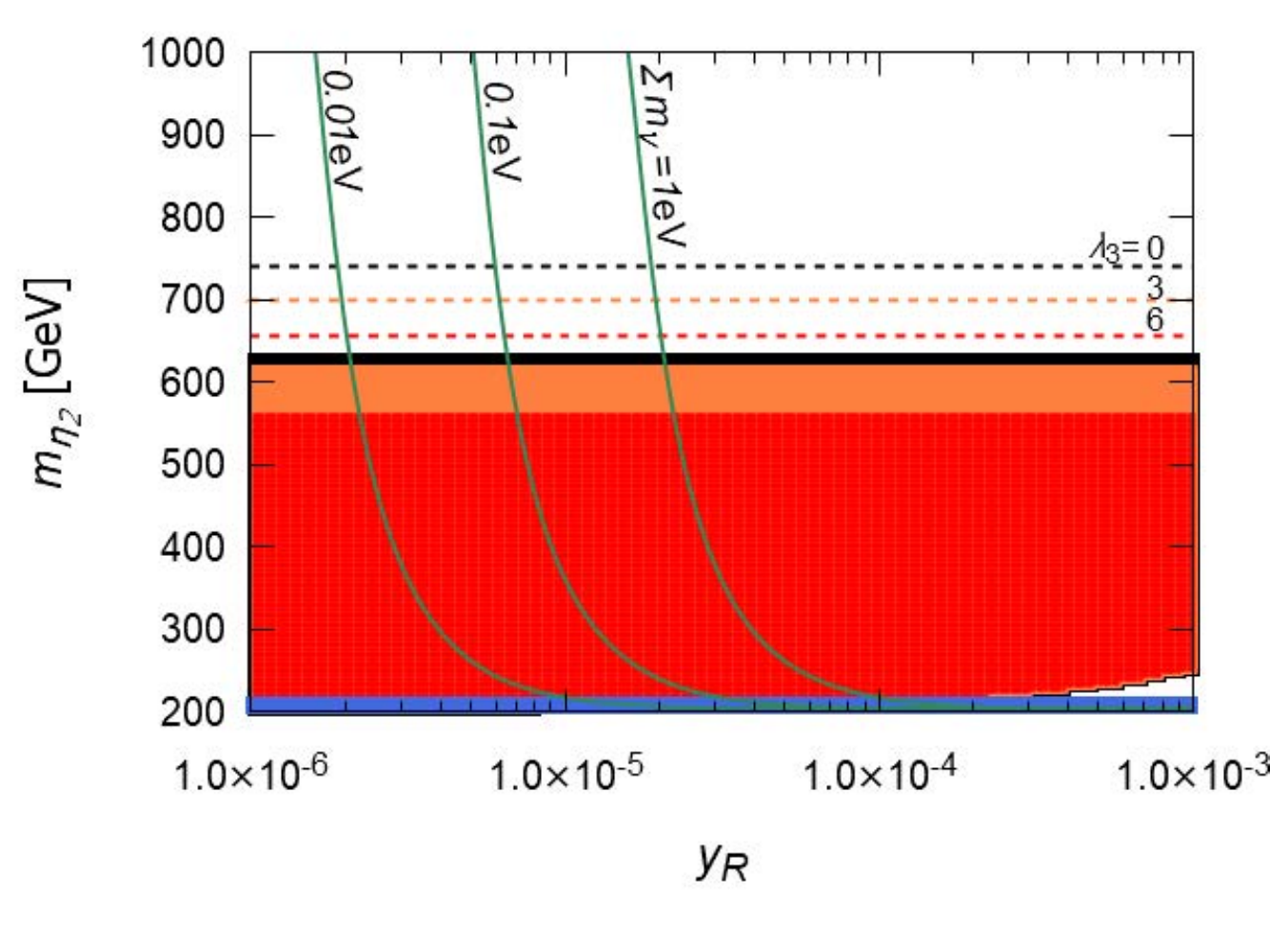} \\
   \caption{Constraints on the parameter space in the $y_R^{}$--$m_{\eta_2^\pm}(=m_{\eta_I})$ plane for $m_{\eta_R}= m_{\eta_1^\pm} =200$~GeV, $\theta = \pi/4$ and $\lambda_{2,6,7,8}=0$.  
We take $M_R = 100$~MeV (upper plots) and 190~GeV (lower plots).  
The efficiency factor $\epsilon$ is taken to be 3\% (left), 30\% (center) and 100\% (right), where the result with $\epsilon=0.3\%$ (value obtained by naive estimation) is the same as that with $\epsilon = 3\%$. 
The regions above the black, orange and red dashed lines are excluded by the perturbative unitarity bound, while the areas shaded in black, orange and red are excluded by the LHC data for $\lambda_3 = 0$, 3 and 6, respectively.  
The region shaded in blue is excluded by the perturbativity bound $(f_L^{\mu}f_R^{\mu} \leq 4\pi)$.  
Each green contour shows the sum of the neutrino masses $\sum_i m_i$. }
   \label{fig:lhc2}
\end{figure}

In Fig.~\ref{fig:lhc2}, we show the constraints on the parameter space in the $y_R$-$m_{\eta_2^\pm}$ plane for 
$m_{\eta_R} = m_{\eta_1^\pm}=200$~GeV and $\theta = \pi/4$.  We take $M_R = 100$~MeV in the upper plots and 190~GeV in the lower plots.  
We also show the prediction of the sum of the neutrino masses given by Eq.~(\ref{eq:ne}) using the green solid curves.  
The efficiency factor $\epsilon$ is taken to be $3\%$, $30\%$ and $100\%$ from left to right panels. 
We see that the constraint from the direct search at the LHC (shaded in red, orange or black) with $\epsilon = 3\%$ vanishes in the region displayed in this figure, 
so that we do not need to show the case with $\epsilon = 0.3\%$ which is the same as that with $\epsilon = 3\%$. 
For $\epsilon = 30\%$ and $100\%$ expected to be obtained by future updates of LHC data, a range of the mass difference $m_{\eta_2^\pm} - m_{\eta_1^\pm}$ is excluded for $y_R < 0.1$
\footnote{For $m_{\eta_2^\pm} = 600$~GeV, $\sigma(pp \to W^{\pm *}\to \eta_2^\pm \eta_I^{})$ and 
$\sigma(pp \to \gamma^*/Z^*\to \eta_2^+ \eta_2^{-})$ are calculated to be about 0.072~fb and 0.037~fb, respectively, while we obtain 
${\cal B}(\eta_2^\pm \to W^\pm \eta_R)\simeq 0.67$, 
${\cal B}(\eta_2^\pm \to Z \eta_1^\pm)\simeq 0.33$, 
${\cal B}(\eta_I^{} \to W^\pm \eta_1^\mp)\simeq 0.50$ and 
${\cal B}(\eta_I^{} \to Z \eta_R)\simeq 0.50$. 
Thus, the cross section of the final state including $W^\pm Z$ is about 0.052~fb, which is slightly larger than the upper limit.  }, 
where a larger $y_R$ and/or a smaller mass difference tends to relax the constraint. This is because 
the decay branching ratios via the gauge coupling in Eq.~(\ref{eq:gauge-decay}) are suppressed by the other decay modes into a fermion pair, i.e., Eq.~(\ref{eq:yukawa-decay}) and/or
by the phase space factor. 
It is also seen that the constraint from the LHC data gets weaker for larger values of $\lambda_3$, because the decay rate $\eta_2^\pm \to \eta_1^\pm h$ becomes larger.
The value of $y_R$ is constrained for a fixed value of $M_R$ and the mass difference $m_{\eta_I}-m_{\eta_R}(=m_{\eta_2^\pm}-m_{\eta_1^\pm})$ by requiring $\sum_i m_{\nu_i} < 0.1$~eV. 
For $M_R = 100$ MeV (upper plots) and 190 GeV (lower plots),  $y_R^{}$ is typically required to be smaller than $1\times 10^{-3}$ and $3\times 10^{-5}$, respectively. 
This behavior can be understood from the expression for the neutrino mass given in Eq.~(\ref{eq:ne}). 


We now comment on the constraint from the slepton searches at the LHC, which is only relevant to Scenario-II.   
As aforementioned, the mass of $\eta_1^\pm$ around 200~GeV has already been excluded by the slepton searches for the case without degenerate $\nu_R$, because of the process given in Eq.~(\ref{eq:slepton}).  
On the other hand, for the case with $M_R \lesssim m_{\eta_R}$ as shown in the right plot of Fig.~\ref{fig:lhc2}, charged leptons produced via the decay of $\eta_1^\pm$ can be too soft to be registered at the detector.  In this case, we can avoid the bound from slepton searches.  Therefore, the successful benchmark scenario with the light $\eta_1^\pm$ is realized in Scenario-II with nearly degenerate $\nu_R$ and $\eta_1^\pm$ in mass.

\subsection{Deviation in the Yukawa coupling from the SM prediction}

\begin{figure}[t]
\begin{center}
 \includegraphics[width=170mm]{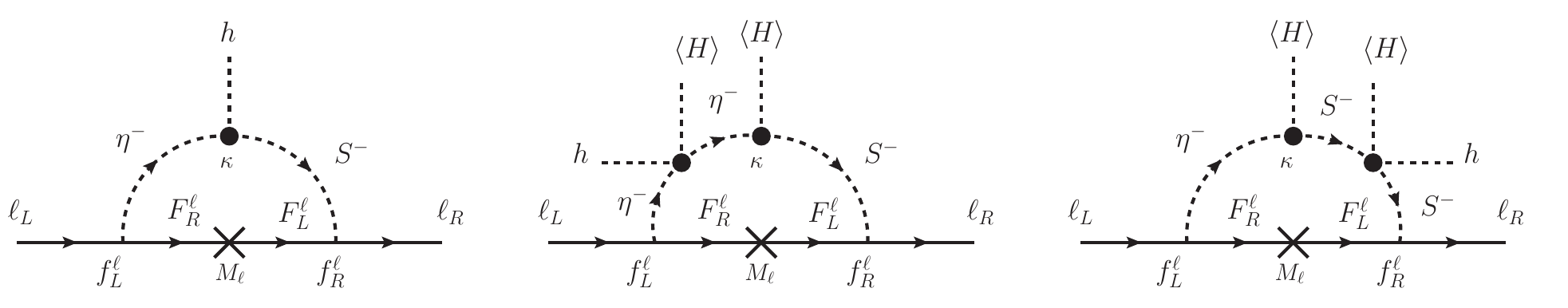}
\caption{One-loop induced Yukawa couplings for charged leptons. }
   \label{fig:yukawa}
\end{center}
\end{figure}

In the SM, Yukawa couplings are determined by $m_f/v$ at tree level.  In our model, this relation for light charged leptons does not hold, because several diagrams different from those of mass generation (see Fig.~\ref{fig:diagram}) contribute to the Yukawa couplings, where scalar quartic couplings enter, as shown in Fig.~\ref{fig:yukawa}. 
The analytic expression for the one-loop induced Yukawa couplings is given by 
\begin{align}
y_\ell =& -\frac{v m_\ell}{F \left(x_1^2 \right)  - F \left(x_2^2 \right)}\Bigg[\sum_{i=1,2}
\sigma_i\lambda_{h\eta_i^+\eta_i^-}C_0(m_{\eta_i^\pm},M_\ell,m_{\eta_i^\pm})  +\frac{\lambda_{h\eta_1^\pm\eta_2^\mp }}{\tan 2\theta} C_0(m_{\eta_1^\pm},M_\ell,m_{\eta_2^\pm})\Bigg]
~,\label{eq:yell}
\end{align}
where $\sigma_i = +1~ (-1)$ for $i = 1~ (2)$ and $C_0$ is the Passarino-Veltman's three-point scalar function~\cite{Passarino:1978jh} 
\begin{align}
C_0(m_1,m_2,m_3) &= -\int_0^1 dx \int_0^1dy\frac{y}{xy[m_1^2-m_2^2 + (y-1)m_h^2] + y(m_2^2 - m_3^2)  + m_3^2}
~. 
\end{align}
Here we have neglected the charged lepton mass in the loop function.  The scalar trilinear couplings $\lambda_{XYZ}$ are defined as the coefficient of the $XYZ$ vertex normalized by the VEV $v$ in the Lagrangian.  We obtain 
\begin{align}
\lambda_{h\eta_1^+\eta_1^-} &= 
    \left( \frac{2m_{\eta_2^\pm}^2}{v^2} -\frac{2m_{\eta_1^\pm}^2}{v^2} \right)c_\theta^2s_\theta^2 - \lambda_3c_\theta^2 - \lambda_7s_\theta^2 
    ~,\\
\lambda_{h\eta_2^+\eta_2^-}& =    -\left( \frac{2m_{\eta_2^\pm}^2}{v^2} -\frac{2m_{\eta_1^\pm}^2}{v^2}  \right)c_\theta^2s_\theta^2- \lambda_7c_\theta^2 - \lambda_3 s_\theta^2
~, \\
\lambda_{h\eta^\pm_1\eta^\mp_2} &=  \frac{s_{2\theta}}{2}\left[\left(\frac{m_{\eta_2^\pm}^2}{v^2} -\frac{m_{\eta_1^\pm}^2}{v^2}\right)c_{2\theta} + \lambda_3 - \lambda_7 \right] \label{eq:lam12h}
~. 
\end{align}
When $\theta = \pi/4$ and $x_{1,2} \ll 1$, the scale factor of the Yukawa coupling, {\it i.e.}, $\kappa_\ell \equiv y_\ell/y_\ell^{\rm SM}$ is approximately expressed as 
\begin{align}
\kappa_\ell \simeq
\frac{1}{x_2^2\ln x_2^2 -x_1^2\ln x_1^2}\left[(x_2^2-x_1^2)(1+\ln x_1x_2) + \frac{v^2}{M_\ell^2}(\lambda_3 + \lambda_7)\ln \frac{x_2}{x_1} \right] 
~.
\label{eq:app}
\end{align}

\begin{figure}[t]
\begin{center}
 \includegraphics[width=100mm]{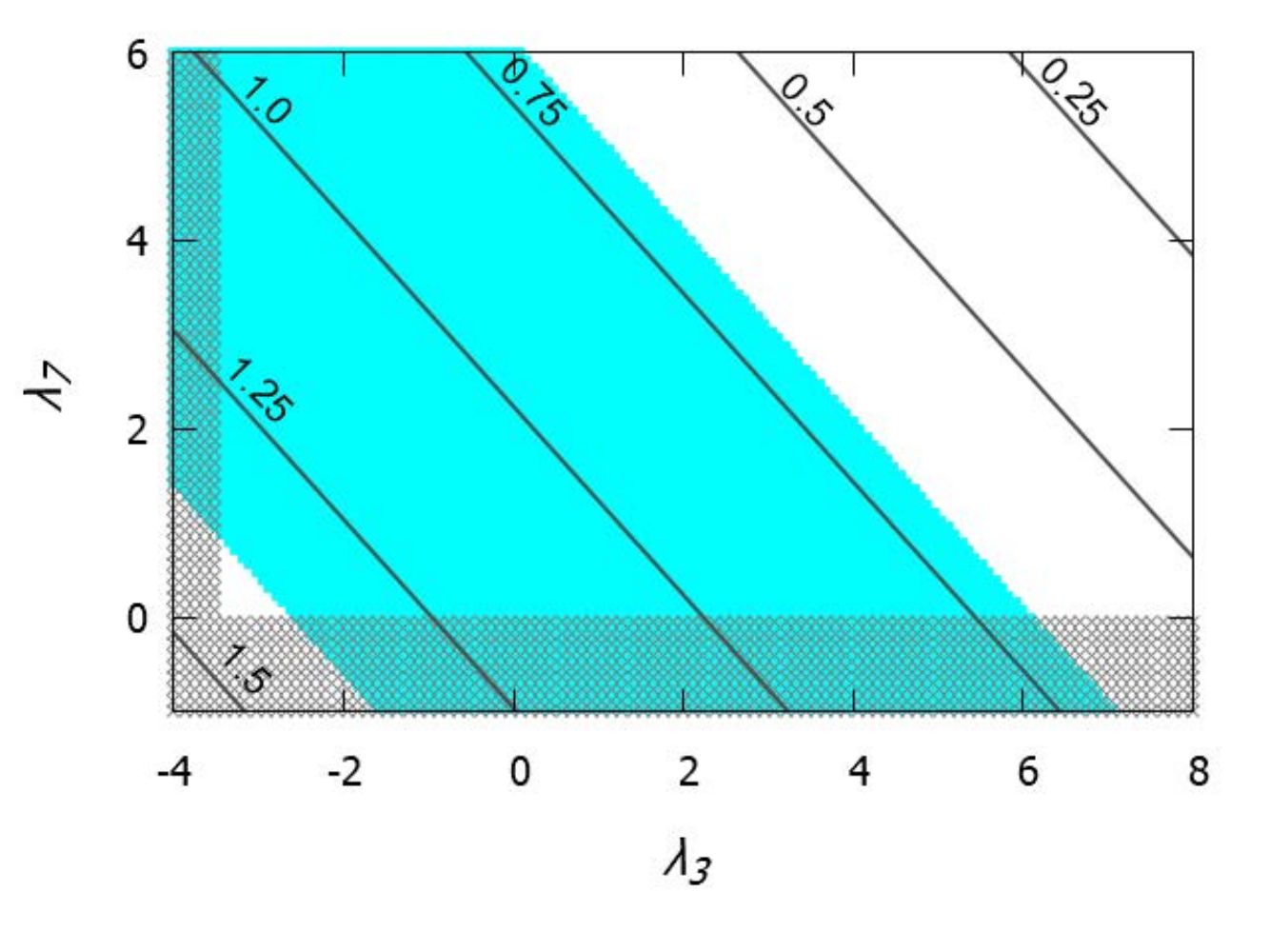}
   \caption{Contour plots for the scale factor $\kappa_\mu$ in the $\lambda_3$-$\lambda_7$ plane for $M_\mu =2$~TeV, $m_{\eta_1^\pm} = 200$~GeV, $m_{\eta_2^\pm} = 600$~GeV and  $\theta = \pi/4$.  The blue shaded region is allowed by the current measurement of the signal strength for the $pp \to h \to \mu\mu$  process at LHC at 95\% CL while the black shaded region is excluded by the bounds from perturbative unitarity and vacuum stability. }
   \label{fig:kappa}
\end{center}
\end{figure}

In Fig.~\ref{fig:kappa}, we show the contour plots of $\kappa_\mu$ for $M_\mu =2$~TeV, $m_{\eta_1^\pm} = 200$~GeV, $m_{\eta_2^\pm} = 600$~GeV and $\theta = \pi/4$, in which the muon $(g-2)$ anomaly can be explained and it is allowed by the bound from the direct searches at LHC.  The blue shaded region is allowed by the current measurements of the signal strength $\mu_\mu$ for the $pp \to h \to \mu\mu$ process at the LHC at $2\sigma$ level, where the weighted average of ATLAS~\cite{Aad:2020xfq} and CMS~\cite{Sirunyan:2020two} is $\mu_{\mu} = 1.19 \pm 0.35$, while the black shaded region is excluded by the constraints from perturbative unitarity and vacuum stability discussed in Sec.~\ref{sec:model}.  We see that larger values of $\lambda_3$ and/or $\lambda_7$ make $\kappa_\mu$ smaller.  This is because the cancellation between the first and second term in Eq.~(\ref{eq:app}) becomes stronger.  We clarify that the value of $\kappa_\mu - 1$ can be $\pm 30\%$ in the parameter region allowed by the constraints.  Such a large deviation in the muon Yukawa coupling can be detected by future collider experiments such as the HL-LHC and the ILC, where the muon Yukawa coupling is expected to be measured with the precision of 7\%~\cite{Cepeda:2019klc} and 5.6\%~\cite{Fujii:2017vwa} at $1\sigma$ level, respectively.  It should be emphasized here that the other Higgs boson couplings, {\it e.g.}, the $hVV$ ($V=W,Z$) and $hf\bar{f}$ ($f\neq e,~\mu$) do not change from the SM predictions at tree level.  Therefore, a large deviation found in the electron and/or muon Yukawa couplings in the future collider experiments could strongly point to our model.

\subsection{Direct searches at $e^+e^-$ colliders}

As shown in the previous subsection, $\eta_1^\pm$ can be of order 100~GeV under the constraints of the direct searches at the LHC.  Such a light charged scalar boson can be directly probed at future lepton colliders. 

In general, any charged particles can be produced in pair from the electron-positron collision via the Drell-Yan process as long as it is kinematically allowed.  In addition, $\eta_1^\pm$ can also be produced via the $t$-channel process with the vector-like lepton $F^e$ exchange in the model.  The cross section for the $e^+e^- \to \eta_1^+ \eta_1^-$ process is then expressed as 
\begin{align}
\sigma_{\rm tot} = \sigma_{s} + \sigma_{t} + \sigma_{st}
~, 
\end{align}
where $\sigma_{s}$, $\sigma_{t}$ and $\sigma_{st}$ represent respectively the contributions from the $s$-channel $\gamma^*/Z^*$ exchange, the $t$-channel $F^e$ exchange and their interference.  Each term is analytically expressed as 
\begin{align}
\sigma_s& = \frac{\beta^3}{48 \pi s} \left[
e^2\vec{c}_\gamma + \frac{g_Z^2}{1-r_Z}\left(\frac{c_\theta^2}{2}-s_W^2\right)\vec{c}_Z \right]^2
~, \\
\sigma_t& = \frac{1}{128\pi s}\left[\frac{r_F\beta (v_F^2-a_F^2)}{(r_1-r_F^{})^2 + r_F^{}} -  2(\vec{c}_F)^2 \beta 
- (\vec{c}_F)^2(1-2r_1+2r_F)\ln \frac{(1 - \beta)^2 + 4r_F^{}}{(1 + \beta)^2 + 4r_F^{}}\right]
~, \\
\sigma_{st}& = \frac{1}{16\pi s}\left[\left(r_F-r_1+\frac{1}{2}\right)\beta + [(r_1-r_F)^2+r_F]\ln \frac{(1 - \beta)^2 + 4r_F^{}}{(1 + \beta)^2 + 4r_F^{}}\right]\notag\\
& \quad \quad\times \left[e^2\vec{v}_\gamma + \frac{g_Z^2}{1-r_Z}\left(\frac{c_\theta^2}{2}-s_W^2\right)\vec{v}_Z  \right]\cdot \vec{c}_F
~, 
\end{align}
where $g_Z^{} = g/c_W$, $s_W^{}$ ($c_W^{}$) is the sine (cosine) of the weak mixing angle, and 
\begin{align}
\vec{c}_\gamma &\equiv (v_\gamma^{},a_\gamma^{})~~\text{with}~~v_\gamma^{} = -1,~a_\gamma^{} = 0, \\ 
\vec{c}_Z &\equiv (v_Z^{},a_Z^{})~~\text{with}~~v_Z^{} = s_W^2-\frac{1}{4},~a_Z^{} = -\frac{1}{4}, \\
\vec{c}_F &\equiv (v_F,a_F)~~\text{with}~~v_F = (f_L^e)^2 c_\theta^2 + (f_R^e)^2 s_\theta^2,~~a_F = (f_L^e)^2 c_\theta^2 - (f_R^e)^2 s_\theta^2. 
\end{align}
We introduce the kinematical variable $\beta \equiv \sqrt{1 - 4r_1}$ with $r_1= m_{\eta_1}^2/s$, $r_F^{} = M_e^2/s$ and $r_Z = m_Z^2/s$.

We see that $\sigma_t$ grows with the fourth power of $f_L^e$ and/or $f_R^e$, so that the cross section is strongly enhanced at large $f_L^e$ and/or $f_R^e$.  We note that it is the combination $f_L^ef_R^es_{2\theta}$ rather than individual sizes of $f_L^e$ and $f_R^e$ that is constrained by the charged lepton masses given in Eq.~(\ref{eq:ml_exact}).  In addition, the term $\sigma_{st}$ typically gives a destructive interference and is more significant than $\sigma_t$, so that for marginal values of $f_{L,R}^e$ the cross section becomes smaller than the case where only $\sigma_s$ contributes to the pair production process.  Therefore, the pair production cross section in our model is generally different from that in models with charged scalar bosons which are produced only via the gauge interactions.  A typical example is the inert doublet model, in which the pair production cross section for the charged scalar bosons is given by $\sigma_s$ with $\theta = 0$.

\begin{figure}[t]
\begin{center}
 \includegraphics[width=80mm]{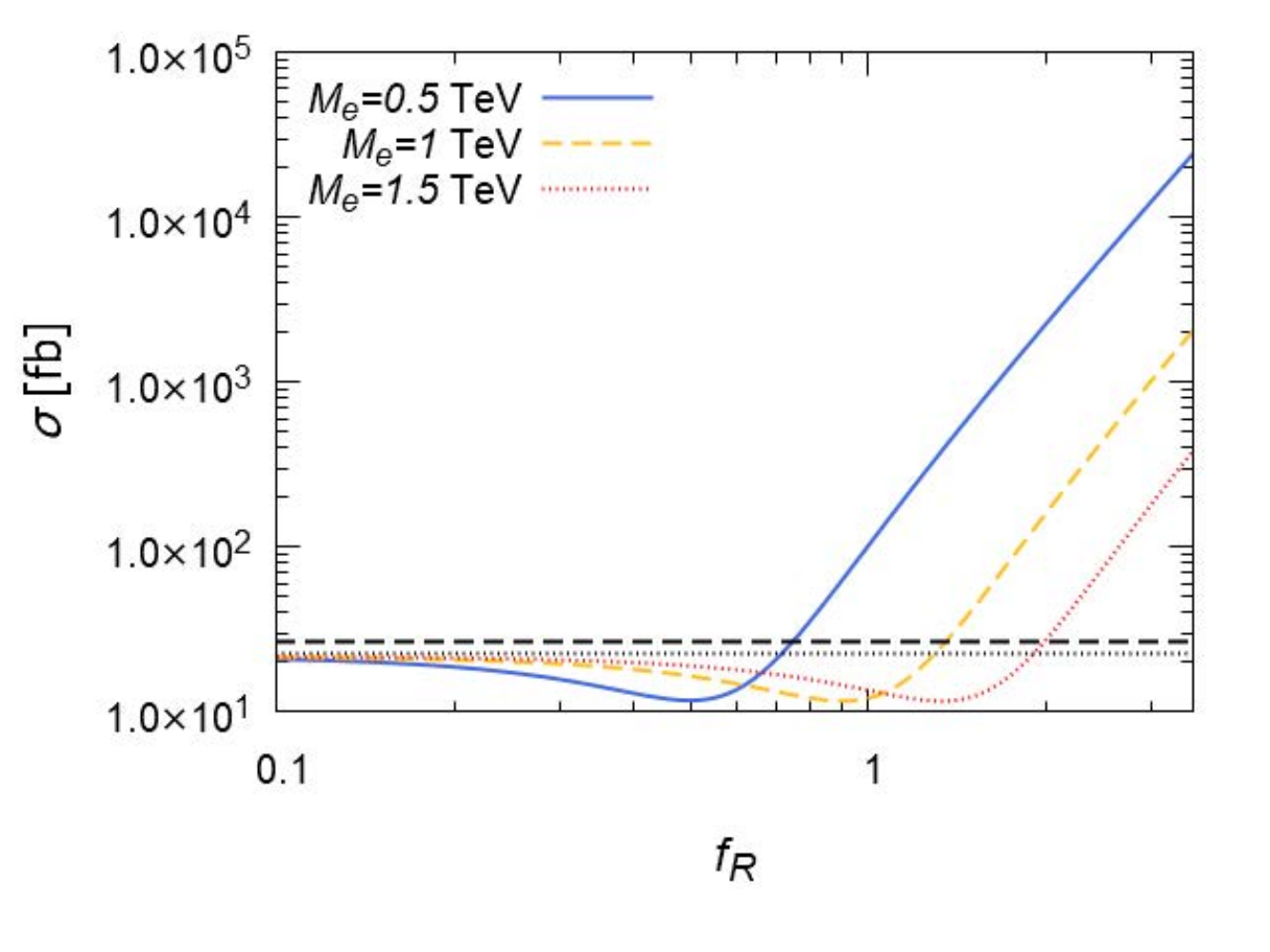}
 \includegraphics[width=80mm]{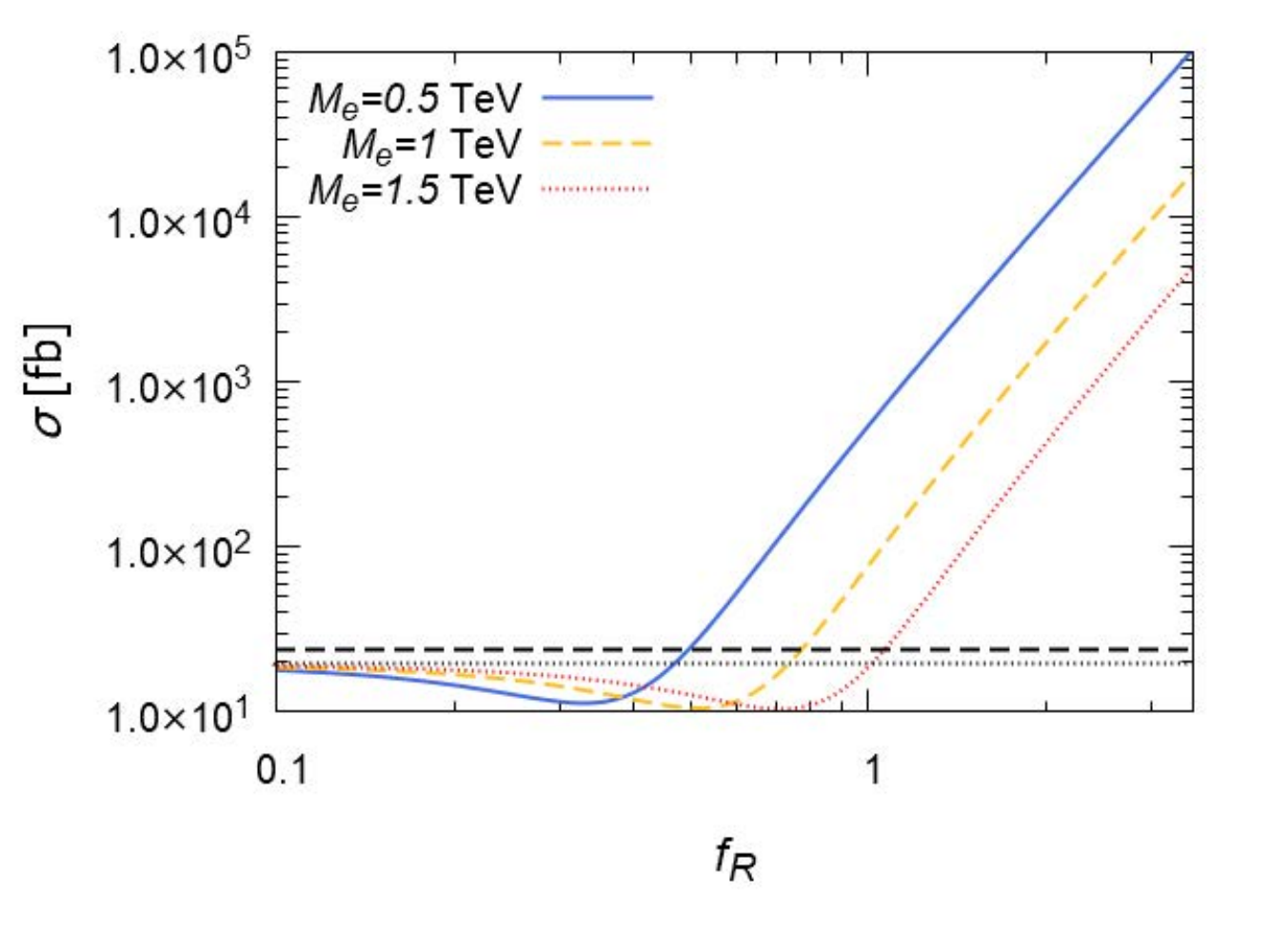}
   \caption{Production cross section for the $e^+e^- \to \eta_1^+\eta_1^-$ process as a function of the Yukawa coupling $f_R^e$ for $m_{\eta_1^\pm} = 200$~GeV and $\theta=\pi/4$.  The left (right) plot shows the case with $\sqrt{s} = 500$ (1000)~GeV.  The solid, dashed and dotted curves show the cases with $M_e=0.5$, 1 and 1.5~TeV, respectively.  The horizontal dashed and dotted lines show the cross section for $\theta = 0$ ($\eta_1^\pm$ corresponding to the pure doublet state) and $\theta = \pi/2$ ($\eta_1^\pm$ corresponding to the pure singlet state), respectively. }
   \label{fig:ilc}
\end{center}
\end{figure}

In Fig.~\ref{fig:ilc}, we show the cross section for the $e^+e^- \to \eta_1^+\eta_1^-$ process as a function of the new Yukawa coupling $f_R^e$ with $m_{\eta_1^\pm} = 200$~GeV and $\theta = \pi/4$.  
The vector-like lepton mass $M_e$ is taken to be 500~GeV (solid), 1~TeV (dashed) and 1.5~TeV (dotted).  The value of $f_L^e$ is fixed by using the formula for the electron mass in Eq.~(\ref{eq:ml_exact}).  We take the center-of-mass energy $\sqrt{s} = 500$ (1000)~GeV in the left (right) plot.  Take the left plot as an example, the cross section has a minimum at $f_R^e = 0.5$, 0.9 and 1.1 for $M_e=500$~GeV, 1~TeV and 1.5~TeV, respectively, and it then becomes larger for larger values of $f_R^e$ due to the enhancement of the $\sigma_t$ contribution.

A careful background and systematic error analysis is necessary in order to verify the feasibility of the detecting $\eta_1^\pm$ at the lepton colliders, which is beyond the scope of this paper.

\section{Conclusions \label{sec:conclusions}} 

We have studied a model with radiative generation of masses for electron and muon as well as left-handed neutrinos at one-loop level.  
Such loop-induced masses are realized by introducing the dark sector, realized by imposing an exact $Z_2$ symmetry, composed of vector-like leptons $F^\ell$, right-handed neutrinos $\nu_R^\ell$ and dark scalars $\eta$ and $S^\pm$.  The lightest neutral particle in the dark sector can be a dark matter candidate.  In this scenario, new contributions to the muon $(g-2)$ anomaly are almost determined only by the mass of vector-like leptons $F^\mu$, because the same particles run in the loop in both the diagrams for the muon mass generation and those for the muon $(g-2)$.  In fact, the muon $(g-2)$ anomaly can be accommodated by taking the mass of $F^\mu$ to be about 2~TeV.  On the other hand, masses and mixings of left-handed neutrinos are generated at one-loop level by right-handed neutrino $\nu_R$ loops.  In order to make a phenomenologically acceptable scenario, we set a key assumption that both lepton number and lepton flavor violations appear only through the Majorana mass terms of $\nu_R$.  In this scenario, observed neutrino oscillations are successfully explained without contradiction to the radiative LFV decays of charged leptons. 

We have discussed constraints from direct searches at current LHC data in Scenario-I ($\nu_R$ being heavier than the dark scalars) and in Scenario-II ($\nu_R$ being lighter than the dark scalars).  
We have shown that a portion of the parameter space in Scenario-I is constrained by the search for chargino-neutralino pair productions, 
whose data can be recast to restrict our model as the $\eta_2^\pm$ and $\eta_I^{}$ production provides similar final states, {\it i.e.}, $\eta_2^\pm \eta_I \to W^\pm Z \eta_R\eta_R$ with $\eta_R^{}$ being the dark matter.  
We have found that the case for $m_{\eta_1^\pm} = 200$ GeV, the dark matter mass ($m_{\eta_R}$) to be 63 GeV and $\theta = 0$  
is now excluded by the direct search at the LHC, while a non-zero mixing case is allowed for a larger mass of $m_{\eta_2^\pm}$, e.g., $m_{\eta_2^\pm} > 500$ GeV (for $\theta \sim 6^\circ $) and $m_{\eta_2^\pm} > 300$ GeV (for $\theta \sim 12^\circ$). 
In Scenario-II, the LHC bound can be relaxed as compared with Scenario-I, because other decay channels of the dark scalar bosons become open, such as $\eta_{R,I} \to \nu_L\nu_R$ and $\eta_{1,2}^\pm \to \ell_L^\pm\nu_R$.
In particular, for the case with $\theta = \pi/4$ and the nearly degenerate masses between $\eta_1^\pm$ and $\nu_R$, 
the current constraints from searches for chargino-neutralino as well as slepton at the LHC do not exclude the parameter space. 
For both the scenarios, the constraint from the direct search at the LHC becomes significantly stronger when the efficiency factor $\epsilon$ gets larger in future updates of LHC data. 
For example, almost all the parameter region in Scenario-I with $m_{\eta_1^\pm} = 200$~GeV and $m_{\eta_R}=63$~GeV can be excluded by taking $\epsilon \gtrsim 30\%$. 

One of the most important predictions in our model is the large deviation in the muon Yukawa coupling with the 125-GeV Higgs boson.  We have shown that the muon Yukawa coupling can deviate at a few tens of percent level in the successful benchmark scenario.  Such a large deviation can easily be probed at the HL-LHC and/or the ILC.  Finally, we have evaluated the cross section for the pair production of the lighter charged scalar boson $\eta_1^\pm$ at electron-positron colliders.  Because of the new Yukawa coupling with electrons, the $t$-channel diagram also contributes to the pair production in addition to the usual Drell-Yan process.  We have found that the cross section can be largely different from models without such a $t$-channel diagram.  In particular, larger values of the cross section of order 100~fb can be obtained at $\sqrt{s} = 500$~GeV even when the vector-like lepton mass is taken to be 1~TeV. 

\vspace*{4mm}

\begin{acknowledgments}
We would like to thank Prof. Shinya Kanemura and Prof. Ryosuke Sato for fruitful discussions.  We also would be grateful to Prof. Kazuki Sakurai for useful comments on the collider phenomenology.  The works of CWC and KY were supported in part by Grant No.~MOST-108-2112-M-002-005-MY3 and the Grant-in-Aid for Early-Career Scientists, No.~19K14714, respectively. 
\end{acknowledgments}

\bibliography{references}

\end{document}